\documentclass[fleqn,usenatbib]{mnras}
\usepackage{graphicx}
\usepackage{natbib}
\usepackage{epstopdf}
\usepackage{amssymb}
\usepackage{amsmath}
\usepackage{mathptmx}
\usepackage{booktabs}
\usepackage{hyperref}
\usepackage{float}
\usepackage{subfig}

\newcommand{\msun}{\,{\rm M_{\odot}}}
\newcommand{\mx}{\,{\rm max}}
\newcommand{\cm}{\,{\rm cm}}
\newcommand{\s}{\,{\rm s}}	
\newcommand{\erg}{\,{\rm erg}}
\newcommand{\rad}{\,{\rm rad}}
\newcommand{\Eiso}{E_{\rm{iso}}}

\newcommand{\Liso}{L_{\rm{iso}}}

\newcommand{\A}{{\it Lc}}
\newcommand{\B}{{\it Lw}}
\newcommand{\C}{{\it Ln}}
\newcommand{\D}{{\it Lp}}
\newcommand{\E}{{\it Lsd}}
\newcommand{\F}{{\it Lnp}}
\newcommand{\G}{{\it Lvp}}
\newcommand{\h}{{\it Llh}}
\newcommand{\I}{{\it Lvwlh}}
\newcommand{\J}{{\it Lvw}}
\newcommand{\SG}{{\it S}}
\usepackage[normalem]{ulem}
\usepackage{xcolor}
\definecolor{new_color}{rgb}{1.0, 0.43, 0.3}

\title[The structure of hydrodynamic GRB jets]{The structure of hydrodynamic $ \gamma $-ray burst jets}
\author{
	Ore Gottlieb \altaffilmark{1}, Ehud Nakar \altaffilmark{1}, Omer Bromberg \altaffilmark{1}, 
}
\author[Gottlieb, Nakar \& Bromberg]{
	Ore Gottlieb\thanks{oregottlieb@mail.tau.ac.il},
	Ehud Nakar,
	Omer Bromberg
	\\
	{School of Physics and Astronomy, Tel Aviv University, Tel Aviv 69978, Israel}
}
\pubyear{2020}
\begin{document}
	\label{firstpage}
	\pagerange{\pageref{firstpage}--\pageref{lastpage}}
	\maketitle	
	\begin{abstract}
		After being launched, GRB jets propagate through dense media prior to their breakout. The jet-medium interaction results in the formation of a complex structured outflow, often referred to as a “structured jet”. The underlying physics of the jet-medium interaction that sets the post-breakout jet morphology has never been explored systematically. Here we use a suite of 3D simulations to follow the evolution of {\it hydrodynamic} long and short gamma-ray bursts (GRBs) jets after breakout to study the post-breakout structure induced by the interaction. Our simulations feature Rayleigh-Taylor fingers that grow from the cocoon into the jet, mix cocoon with jet material and destabilize the jet. The mixing gives rise to a previously unidentified region sheathing the jet from the cocoon, which we denote the {\it jet-cocoon interface} (JCI). lGRBs undergo strong mixing, resulting in most of the jet energy to drift into the JCI, while in sGRBs weaker mixing is possible, leading to a comparable amount of energy in the two components. Remarkably, the jet structure (jet-core plus JCI) can be characterized by simple universal angular power-law distributions, with power-law indices that depend solely on the mixing level. This result supports the commonly used power-law angular distribution, and disfavors Gaussian jets. At larger angles, where the cocoon dominates, the structure is more complex. The mixing shapes the prompt emission lightcurve and implies that typical lGRB afterglows are different from those of sGRBs. Our predictions can be used to infer jet characteristics from prompt and afterglow observations.
	\end{abstract}
	
	\section{Introduction}
	\label{sec_introduction}
	
	A Gamma-Ray Burst (GRB) arises after an ultra relativistic collimated jet breaks out from a dense medium, be it a star in long GRBs (lGRBs) or, presumably, a Neutron star merger ejecta in short GRBs (sGRBs). 
	In both cases the jet has to successfully cross the circumventing medium before it can form a GRB. As it propagates through the medium the jet drives a bow shock ahead of it, behind which a cocoon of shocked matter is formed. The cocoon applies pressure on the jet, collimates it and facilitates its passage through the medium. The jet, and the cocoon that engulfs it, interact with each other as they co-evolve. This interaction plays a crucial role in shaping the morphology of the jet and the cocoon, both inside the dense medium and after breaking out of it.
	
	The post-breakout structure ultimately governs the observational imprint of the jet from the prompt emission to the afterglow, thus, characterizing it is of a great interest.
	In most models that can be found in the literature the jet structure is assumed to be top-hat. Namely, the jet is considered to have uniform distributions of Lorentz factor and power within a given opening angle and zero outside. While such a model could be sometimes satisfactory for jets that are observed from directions within the jet cone, it is highly unlikely to be sufficient for ``off-axis" observers located at large viewing angles from the axis. 
	Recent observations of GRB170817 indicated that the jet was viewed at an offset of $ \sim 20^\circ $ from the axis \citep{Mooley2018b}. Based on these and other afterglow observations it was confirmed that jets are likely to have a more complex structure, often referred to as ``structured jet" \citep[e.g.][see \citealt{Nakar2019} for a review of GW170817 and more references therein]{Alexander2018,Dobie2018,Fong2019,Ghirlanda2019,Hajela2019,Mooley2018a,Mooley2018b,Mooley2018c,Troja2019,Wu2019}. This structure has a profound effect on the observed emission at angles larger than the jet opening angle.
	
	The structure of the jet is often assumed to have a characteristic angular dependent structure (e.g. proper-velocity and energy distributions) that can be described analytically. The most commonly used functions are Gaussian and a power-law with a core functions (e.g. \citealt{Lipunov2001,Rossi2002,Kumar2003,Rossi2004,Lamb2017,Gill2018,Lyman2018,Margutti2018,Resmi2018,Troja2018,Xie2018,Lamb2019,Meng2019,Oganesyan2019}). Another form commonly used is a boosted fireball \citep{Duffell2013b}. Such parameterized functions are not motivated by any physical mechanism, and their parameters are typically set by fitting to observations.
	
	An alternative approach to model the jet structure is to study the jet formation and evolution inside the medium as well as after it breaks out. There are at least two major sites that control the final jet structure. (i) The launching site, close to the central compact object. This is where the jet is first accelerated and collimated. There are attempts of modeling the jet structure upon emergence from near the compact object \citep[e.g.,][]{Fernandez2019,Kathirgamaraju2019}. This is a challenging task since our understanding of the launching process is still limited and the outcome depends on unknown initial parameters such as the magnetic field structure in the disk. These studies also ignore the effects of the medium on the jet evolution and on its final structure. Here we do not address this phase.
	(ii) The medium surrounding the compact object, through which the jet propagates. Jet-medium interaction plays a major role in shaping the jet structure. The process was studied analytically \citep[e.g.,][]{Bromberg2011b,Nakar2017,Lazzati2019,Salafia2019} and numerically \citep[e.g.,][]{Ito2015,Kathirgamaraju2017,Xie2018,Lazzati2017c,Gottlieb2018a} in the past, and was found to generate the jet-cocoon. However, these works did not study the details of the jet structure and its dependence on system parameters such as the jet luminosity, the injection angle or the medium density profile, nor did they study the outcome of the interaction between the jet and its cocoon. These topics are at the focus of the current work.
	
	We use numerical simulations  to characterize the post-breakout structure that emerges from the interaction of a jet with the medium it is injected into.
	Previous studies have shown that the structure depends strongly on the mixing, which takes place both inside the cocoon and along the interface between the jet and the cocoon  \citep{Morsony2007,Mizuta2013,LopezCamara2013,LopezCamara2016,Harrison2018}. The mixing in both regions is a result of hydrodynamic instabilities that can be studied only in 3D \citep{Gottlieb2018a}. 3D simulations show that even when the injection of the jet and the medium profile are axisymetric, the instabilities along the interface between the jet and the cocoon break the symmetry and the 3D distribution functions in the jet-cocoon system becomes asymmetric \citep[e.g.,][]{Ito2015,Gottlieb2019b}.
	
	Hydrodynamic (unmagnetized) jets\footnote{For a companion work on the structure of weakly magnetized jets interacting with media see \citet{Gottlieb2020b}} are prone to a variety of instabilities, such as Kelvin-Helmholtz instability (KHI; \citealt{Helmholtz1868,Thomson1871}) and Rayleigh-Taylor instability (RTI; \citealt{Rayleigh1882,Taylor1950}). The former forms when there is a tangential shear velocity between two fluids, as may occur in the interface between the jet and the cocoon (e.g. \citealt{Meliani2007,Rossi2008,Meliani2010}). RTI takes place whenever a lighter fluid accelerates into a heavier one, as in a lateral acceleration of the jet into the cocoon (e.g.  \citealt{Meliani2010,Matsumoto2013a,Matsumoto2013,Matsumoto2017,Matsumoto2019,Toma2017,Gourgouliatos2018}) or at the jet head \citep{Duffell2013,Duffell2014}.
	To date, only a few works \citep{Rossi2008,Meliani2010} have explored the hydrodynamic instabilities in jets with a full 3D setup, none of which has examined its effect on the structure of GRB jets after they break out from a dense medium.
	One immediate profound implication of the mixed nature of the jet is manifested in its prompt emission. \citet{Gottlieb2019b} showed that the mixing in the jet affects the efficiency of the photospheric emission, induces  temporal variations, promotes internal shocks and may account for the rapid variability of the prompt emission.
	
	In this paper we study the jet structure by carrying out a set of 3D simulations with different jet and medium properties, relevant for lGRBs and sGRBs. In these simulations we launch a jet continuously at the center of a dense medium and follow its propagation, the formation of the cocoon, the breakout from the medium, and the post-breakout evolution of the jet-cocoon system. At the end of our simulations most of the outflow expands homologously and stops evolving, so we can study the final structure of the jet-cocoon outflow. Since the goal of the paper is to study the effect of the jet-medium interaction on the emerging structure, and we have no specific information on the jet structure at the launching point, we inject a uniform conical jet (i.e., top-hat), so the entire structure is induced by the interaction. We leave the study of the launching process on the jet structure to a future work. 
	
	We find that although the jet-cocoon system generally has an asymmetric three dimensional structure, it can be approximated reasonably well (at least for the purpose of the afterglow emission) by an axisymetric distribution with a well defined angular dependent proper-velocity. 
	Our main finding is that the final structure of the system can be divided into three components: (i) The jet (ii) The cocoon (iii) A  mixed jet-cocoon material between them that emerges from the continuous mixing between the jet and the cocoon along their contact surface. We denote this region as the ``jet-cocoon interface" (JCI).  The first two components, the jet and the cocoon, were discussed in the past (e.g., \citealt{Nakar2017,Lazzati2017c}). However, although the JCI was present in previous simulations (e.g., \citealt{Lazzati2017c,Gottlieb2018a}), it was never recognized as a distinct component, and it is explored here in detail for the first time.
	
	The outline of this paper is as follows. In \S\ref{sec:models} we set up the numerical framework and present a range of lGRB and sGRB models that we consider. In \S\ref{sec:evolution} we discuss the structure of the jet and the cocoon before breakout while the jet head is still inside the dense medium. In \S\ref{sec:structure} we characterize the terminal distributions of jet structures after breakout and discuss the similarities and differences between lGRBs and sGRBs as well as between 3D simulations, 2D simulations, and common analytic modeling. Finally, in \S\ref{sec:emission} we discuss how the resulting structure reshapes the expected prompt and afterglow light curves before concluding  in \S\ref{sec:discussion}.
	
	\section{Models}
	\label{sec:models}
	\begin{table*}
		\setlength{\tabcolsep}{7.5pt}
		\centering
		\begin{tabular}{ | l | c  c  c  c  c  c  c | }
			
			\hline
			lGRB Model & $ L_j [10^{50}\rm{erg~s^{-1}}] $ & $ \theta_{j,0} = 0.7\Gamma_0^{-1} $ & $ u_{\infty,\mx} $ & $ M_\star [\msun] $ & $ \rho_\star(r) $ & $ t_b [\rm{s}] $ & $ t_e [\rm{s}] $\\ \hline
			$ \A^{\dagger 1} $ (canonical) & $ 1.0 $ & $ 0.14 $ & $ 500 $ & $ 10 $ & $ \rho_0(r/r_0)^{-2}x^3 $ & $ 20 $ & 68\\
			$ \B $ (wide) & $ 1.0 $ & $ 0.18 $ & $ 400 $ & $ 10 $ & $ \rho_0(r/r_0)^{-2}x^3 $ & $ 23 $ & 41\\
			$ \C $ (narrow) & $ 1.0 $ & $ 0.07 $ & $ 1000 $ & $ 10 $ & $ \rho_0(r/r_0)^{-2}x^3 $ & $ 13 $ & 24\\
			$ \D $ (powerful) & $ 5.0 $ & $ 0.14 $ & $ 500 $ & $ 10 $ & $ \rho_0(r/r_0)^{-2}x^3 $ & $ 8 $ & 36 \\
			$ \E $ (steep $ \rho $ profile) & $ 1.0 $ & $ 0.14 $ & $ 500 $ & $ 2.5 $ & $ \rho_0(r/r_0)^{-2.5}x^3 $ & $ 8 $ & 16\\
			$ \F $ (narrow powerful) & $ 7.0 $ & $ 0.07 $ & $ 1000 $ & $ 10 $ & $ \rho_0(r/r_0)^{-2}x^3 $ & $ 6 $ & 33\\
			$ \G^{\dagger 2} $ (very powerful) & $ 16 $ & $ 0.14 $ & $ 540 $ & $ 10 $ & $ \rho_0(r/r_0)^{-2}x^3 $ & $ 5 $ & 16\\
			$ \h $ (low h) & $ 1.0 $ & $ 0.14 $ & $ 100 $ & $ 10 $ & $ \rho_0(r/r_0)^{-2}x^3 $ & $ 13 $ & 43\\
			$ \I $ (very wide low h) & $ 1.0 $ & $ 0.24 $ & $ 300 $ & $ 10 $ & $ \rho_0(r/r_0)^{-2}x^3 $ & $ 27 $ & 55\\
			$ \J $ (very wide) & $ 1.0 $ & $ 0.24 $ & $ 500 $ & $ 10 $ & $ \rho_0(r/r_0)^{-2}x^3 $ & $ 28 $ & 69\\
			\hline
			sGRB Model & $ L_j [10^{50}\rm{erg~s^{-1}}] $ & $ \theta_{j,0} {= 0.7\Gamma_0^{-1}} $ & $ u_{\infty,\mx} $ & $ M_{ce} [\msun] $ & $ \rho_*(r,\theta)~[\rm{g~cm^{-3}}] $ & $ t_d; t_b [\rm{s}] $ & $ t_e [\rm{s}] $ \\ \hline
			$ \SG_1^{\dagger 3} $ & $ 1.4 $ & $ 0.07 $ & $ 200 $ & $ 0.04  $ & $ 10^{22} (r/\cm)^{-2}\Big(\frac{1}{4}+\rm{sin}^8\theta\Big) $ & $ 0.2; 0.4 $ & $ 1.0 $\\
			$ \SG_2^{\dagger 4} $ & $ 6.7 $ & $ 0.18 $ & $ 100 $ & $ 0.05 $ & $ 5.5\times 10^{34}(r/\cm)^{-3.5} $ & $ 0.7; 1.4 $ & $ 4.9 $\\
			$ \SG_3^{\dagger 5} $ & $ 0.3 $ & $ 0.14 $ & $ 500 $ & $ 0.05 $ & $ 2.2\times 10^{21}(r/\cm)^{-2} $ & $ 0.6; 1.4 $ & $ 3.6 $\\\
			$ \SG_4 $ & $ 10^{-3} $ & $ 0.14 $ & $ 500 $ & $ 0.05 $ & $ 2.2\times 10^{21}(r/\cm)^{-2} $ & $ 0.6; 5.6 $ & $ 11.1 $\\\hline
			
		\end{tabular}
		\hfill\break
		
		\caption{The simulations configurations. $ L_j $ is the total jet luminosity (two sided), $ \theta_{j,0} $ is the jet launching opening angle, $ u_{\infty,\mx} = \sqrt{h_0^2\Gamma_0^2-1} $ is the terminal proper-velocity of the jet, had it not experienced any mixing, is defined by the initial Lorentz factor $ \Gamma_0 $ and the initial specific enthalpy $ h_0 $, $ M_\star $ is the stellar/ejecta mass, $ \rho_\star(r) $ is the radial density profile of the star/ejecta, where $ \rho_0 $ and $ r_0 $ are the density and radius normalizations, respectively, and $ x \equiv (R_\star-r)/R_\star $. $ t_d, t_b, t_e $ are the delay time, breakout time of the forward shock from the dense medium and engine working time, respectively. In the sGRB models all times are measured from the time of the merger and $ t_b $ refers to the breakout from the core ejecta. For models $ \A $ and $ \G $ we also perform axisymmetric 2D simulations for comparison.
			\newline
			$ ^{\dagger 1}$\scriptsize{Models $ \A, \D, \I, \J, \h $ are models $ {\it A, B, C, D, E} $ in \citet{Gottlieb2019b}, respectively.}\\
			$^{\dagger 2}$\scriptsize{\citet{Harrison2018}.}\\
			$^{\dagger 3}$\scriptsize{\citet{Mooley2018b}.}\\
			$^{\dagger 4}$\scriptsize{\citet{Gottlieb2018b}.}\\
			$^{\dagger 5}$\scriptsize{\citet{Gottlieb2019b}.}
		}
		\label{tab_models_comparison}
	\end{table*}

	We examine the evolution of the jet-cocoon system in a collection of setups  expected in lGRBs and sGRBs. For this purpose we carry out 3D relativistic-hydrodynamic (RHD) simulations with a variety of jet powers, opening angles, terminal Lorentz factors and media in which the jets propagate. The full characteristics of all models are listed in Table $ \ref{tab_models_comparison} $.
	For lGRB jets, we use the Collapsar model with a static, non-rotating star of radius $ R_*=10^{11}\cm $, and vary the stellar mass and density profile. 
	We also conduct two 2D lGRB simulations, which are similar to the setups of models $ \A $ and $ \G $, for comparison. We show that 2D models produce results that are considerably different than 3D ones.
	
	For sGRBs we consider ejecta that emerge following a double neutron star (NS) merger \citep[see][for a review]{Nakar2019}. The ejecta was predicted by many theoretical studies and its presence was later confirmed by observations of GW170817. We assume an ejecta mass $\sim0.05\msun$, as inferred from GW170817. The outflow in our simulations is composed of three components:
	(i) A collimated relativistic jet launched from the origin with a delay $ t_d $ after the merger time.
	(ii) A non-relativistic ($ v_c < 0.2 $c, where c is the speed of light) cold core ejecta with a mass $ M_{ce} \approx 0.05 \msun $.
	(iii) A mildly-relativistic cold tail ejecta with a mass $ \sim 0.05  M_{ce} $. Unlike the core ejecta, the tail component has not been directly observed, but has been indicated to be part of these systems by previous studies \citep{Hotokezaka2012,Hotokezaka2018a,Kyutoku2012,Bauswein2013,Beloborodov2018,Radice2018}. We stress that in this work the dilute tail ejecta is expected to have a negligible effect on the mixing and the distributions and thus can be ignored. Nevertheless, this component is important when considering the radiation emitted by the jet and the cocoon at early times. In particular it can be crucial for the shock breakout mechanism to account for the $ \gamma $-ray signal in GW170817 \citep{Gottlieb2018b}. Since we use simulations from previous works (\citealt{Mooley2018b,Gottlieb2018b}, listed as $ \SG_1 $ and $ \SG_2 $, respectively) which focused on modeling the electromagnetic signals in GW170817, this component is included in our models.
	Both components of the ejecta expand homologously and in general can have density profiles that depend on both the radial and the angular coordinates. That is, the ejecta velocity profile is $ v(r,t) = r/t $.
	We provide the main models' characteristics in Table \ref{tab_models_comparison}.
	
	All simulations have been carried out with \textsc{pluto} v4.2 \citep{Mignone2007}, using a relativistic ideal gas equation of state.
	Our integration setup includes a third order Runge-Kutta time stepping, piece-wise parabolic reconstruction with harmonic limiter, and an HLL Riemann solver.
	We use a Cartesian grid where the jet is injected along the $\hat{z}$ axis from the center of the lower boundary.
	The jet engine operates throughout the entire time of the simulations.
	We inject an axisymmetric cylindrical flow with velocity and energy profiles scaling as $ 1/\rm{cosh}\big(\frac{r}{r_{\rm{noz}}}\big)^{8} $, where $ r $ is the cylindrical radius coordinate and $ r_{\rm{noz}} = 10^8\cm $ is the typical nozzle radius.
	The jet material is initially relativistically hot and it is launched with an initial Lorentz factor $ \Gamma_0$. It expands sideways soon after the injection and assumes a conical shape with a half-opening angle\footnote{We verify that injecting the jet conically produces similar results.} $ \theta_{j,0} = 0.7/\Gamma_0 $ \citep{Mizuta2013,Harrison2018}, and thus we inject it at height $ z_{\rm{beg}} = r_{\rm{noz}}/\theta_{j,0} $, where the origin is the center of the progenitor.
	We find that as long as $ z_{\rm{beg}} \lesssim 10^{-2}R_\star $, the evolution of the system is not affected by the injection height or the corresponding nozzle radius (see Appendix \ref{sec:convergence} for convergence tests for the nozzle size). Therefore our nozzle is kept fixed in all simulations.
	
	The 3D lGRB simulation grids are identical to each other (except for simulation $ \G $ which was carried out in \citealt{Harrison2018}). 
	The grid is divided into three patches along the $\hat{x}$ and $\hat{y}$ axes independently, and two patches along the $\hat{z}$-axis. The inner $ x $ and $ y $ axes cover the inner $ |5\times 10^8\cm| $ with 50 uniform cells. The outer patches are stretched logarithmically to $ |3\times 10^{11}\cm| $ with 150 cells on each side. The $ z $-axis has one uniform patch inside the star from $ z_{\rm{beg}} $ to $ R_\star $ with 800 cells, and another logarithmic patch with 1200 cells up to $ 10R_\star $. The total number of cells is therefore $ 350\times 350\times 2000 $. In Appendix \ref{sec:convergence} we verify that we reach convergence with this resolution.
	Simulation $ \G$  was carried out with a higher resolution, twice as many cells on the $ x $ and $ y $ axes and an increase by $ 20\% $ on the $ z $-axis.
	
	The 3D sGRB simulations' grids are different from each other since they are performed as parts of different studies. The full grid setups of $ \SG_1 $ and $ \SG_2 $ are described in \citet{Mooley2018b,Gottlieb2018b}, respectively. The setup of $ \SG_3 $ and $ \SG_4 $ is as follows. We use three patches along the $ x $ and $ y $ axes independently. The inner patch is uniform inside $ |10^9|\cm $ with 160 cells, the outer patches stretch to $ |10^{11}\cm| $ with 400 logarithmic cells in each. Along the $ z $-axis we have 1500 uniform cells until $ 1.2\times 10^{11}\cm $. In total we have $ 960 \times 960 \times 1500 $ cells.
	
	The 2D simulations are conducted in a cylindrical grid with two patches on each axis. On $ r $-axis, one uniform patch with 400 cells to $ 2.5\times 10^9\cm $, and an outer logarithmic patch with 600 cells to $ 3\times 10^{10}\cm $. On the $ z $-axis we employ 1000 uniform cells from $ z_{\rm{beg}} $ to $ 10^{11}\cm $ followed by 2000 logarithmic cells to $ 10^{12}\cm $. In total each 2D grid includes $ 1000 \times 3000 $ cells.
	
	\section{Jet structure inside a dense medium: Evolution \& mixing}
	\label{sec:evolution}
	
	In this section we discuss the structure of the jet and the cocoon while the jet head is propagating in the dense medium. This will help us later to understand the jet-cocoon structure after the jet breaks out, which is the main interest of this paper. We first describe the overall hydrodynamics of the jet-cocoon system, and then discuss the origin of the instabilities at the jet boundaries and how the induced mixing affects the evolution and the structure of the jet.
	
	\begin{figure*}
		\centering
		\includegraphics[scale=0.375, angle=90]{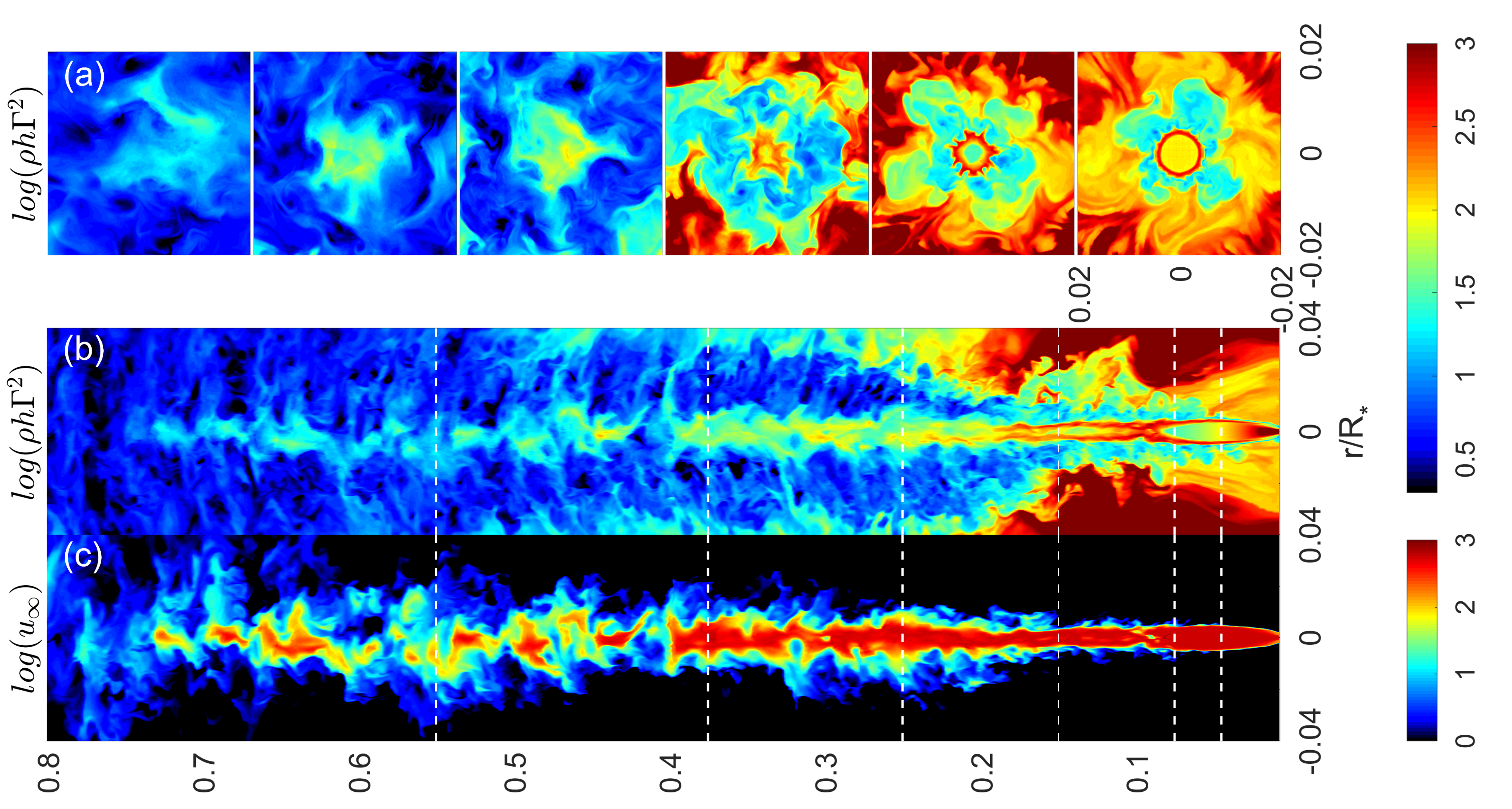}
		\includegraphics[scale=0.25, trim=7cm 0cm 0cm 0cm, angle=0]{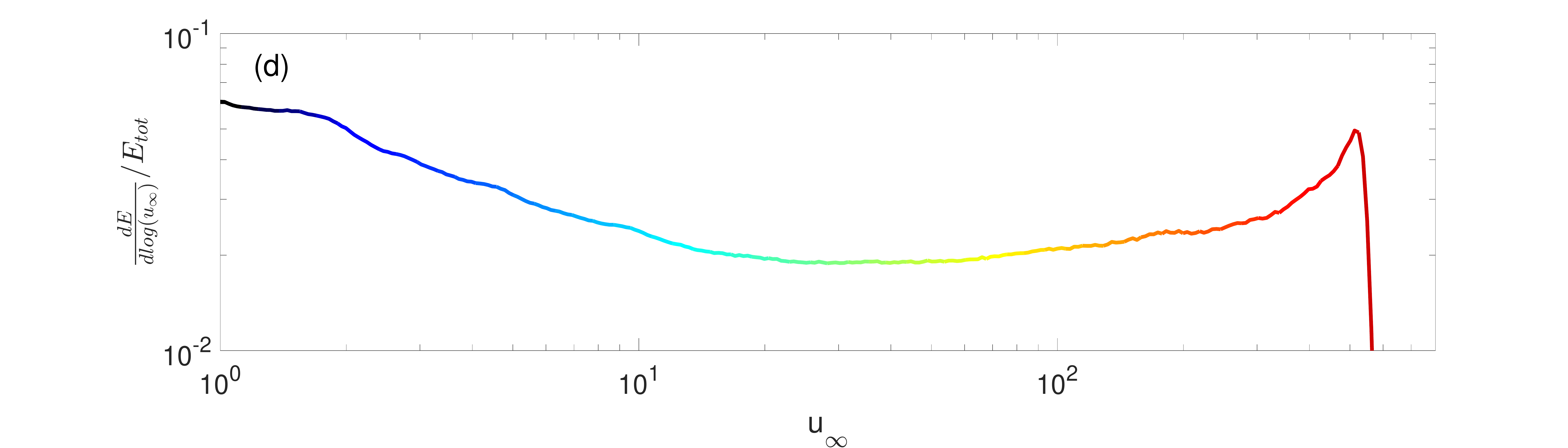}
		\caption[Hydro jet cuts]{
			The jet evolution in model $ \A $ upon breakout. (a) Cuts on the $ x-y $ plane (perpendicular to the jet) of the logarithmic enthalpy density of the jet, $ \rho h\Gamma^2 $. Middle panels show the jet on the $ x-z $ plane (parallel to the jet axis) of the logarithmic enthalpy density of the jet, $ \rho h\Gamma^2 $ (b), and the terminal proper-velocity of an element, had it not underwent any further mixing (c). The dashed white lines represent the places in which the head-on inertia plane cuts are taken. (d) The energy distribution above the collimation shock, normalized by the total energy (excluding the rest-mass energy) at this time. The colors show the mixing of the different elements, in correspondence to the proper-velocity map.
			Videos of the simulation hydrodynamics and instabilities are available at \url{http://www.astro.tau.ac.il/~ore/instabilities.html}.
		}
		\label{fig:hydro_Cuts}
	\end{figure*}

	During the propagation through the medium the jet-cocoon system can be divided into four parts: i) the jet; ii) the jet head, which mediates the jet energy into the cocoon; iii) the inner cocoon, composed of shocked jet material spilled from the jet head and iv) the outer cocoon, composed of shocked medium material that crossed the bow shock. 
	The former three parts are seen in Figure \ref{fig:hydro_Cuts} showing a lGRB from model $ \A $ just before breakout.
	
	Top six panels (\ref{fig:hydro_Cuts}a) show cross sectional cuts of the jet at different altitudes. The color scheme shows the logarithm of the enthalpy density in the lab frame, $ \rho h\Gamma^2 $.
	The two middle panels (\ref{fig:hydro_Cuts}b, \ref{fig:hydro_Cuts}c) show meridian cuts of the jet on the $x-z$ plane. The white dashed lines mark the locations of the cross sectional cults in panels (\ref{fig:hydro_Cuts}a). Color schemes depict (in log scale), (\ref{fig:hydro_Cuts}b) the enthalpy density, and (\ref{fig:hydro_Cuts}c) the asymptotic proper-velocity, $u_\infty \equiv (\Gamma\beta)_\infty= \sqrt{h^2\Gamma^2-1} $, assuming: no further mixing takes place at later stages, and no work is done by the fluid (e.g. through adiabatic expansion). Both of these effects might reduce the terminal proper-velocity of the fluid.
	The collimation of the jet is done through a series of collimation shocks, the first two are seen at $z\leq0.2 R_*$. The unshocked jet material upstream of the first collimation shock is seen in panels (\ref{fig:hydro_Cuts}b) and (\ref{fig:hydro_Cuts}c) at $z\leq0.1 R_*$ and in the two right most cross sectional cuts (\ref{fig:hydro_Cuts}a) as yellow and green circles at the middle of the jet. Instabilities begin to grow on the contact discontinuity between the jet and the cocoon at $z\gtrsim0.05 R_*$, where the first collimation shock begins to converge to the axis. The instabilities continue to grow, mix the jet and cocoon material on the JCI, and eventually destroy the ordered structure of the jet above the second collimation shock at $z \gtrsim 0.2 R_*$. Last, the inner and outer cocoons are seen in panels (\ref{fig:hydro_Cuts}a) and (\ref{fig:hydro_Cuts}b) as light blue-orange and red color regions respectively. The boundary between the two parts is unstable as well, resulting in some mixing between them. The bottom panel (\ref{fig:hydro_Cuts}d) shows the distribution of the total energy  excluding the rest-mass energy, $ E \equiv \int (T_{00}-\rho\Gamma) dV$, in the box per logarithmic unit of $u_\infty$, a measure for the amount of mixing in the jet. We exclude in this analysis the contribution from the unshocked jet material in the collimation shock. The color scheme are the same as in Figure (\ref{fig:hydro_Cuts}c). The unmixed jet material occupies the range of $u_\infty\gtrsim 100$, and the cocoon $u_\infty\lesssim 3$. Everything in between is mixed jet-cocoon material (the JCI).
	
	The growth of instabilities on the jet-cocoon boundary is controlled by the properties of the jet and the cocoon and it is sensitive to the dimensionless parameter $\tilde{L}_c$, which measures the ratio of enthalpy densities between the jet and the cocoon on the jet boundary, defined as \citep{Matsumoto2017}
	\begin{equation}\label{eq:cocoon_ratio}
	\tilde{L}_c = \frac{\rho_jh_j\Gamma_j^2}{\rho_ch_c\Gamma_c^2}~,
	\end{equation}
	where quantities with subscripts $ j $ and $ c $ represent the unshocked jet and inner cocoon material, respectively.
	Previous studies found that whenever $ \tilde{L}_c>1 $ the jet boundary becomes unstable  \citep{Meliani2009,Matsumoto2013,Matsumoto2017,Matsumoto2019}.
	In all of our GRB setups $ \tilde{L}_c > 1$, namely the effective relativistic mass of the jet is larger than that of the cocoon. At the same time, above the collimation point the pressure in the cocoon is  larger than that in the jet, resulting in the jet collimation. 
	This resembles a situation of a ``heavy" fluid (the jet) laying on top of a ``light" fluid (the cocoon) in a gravitational field pointing downwards (outside). Such a condition is unstable for Rayleigh-Taylor instability (RTI) that grows when the ``light" cocoon material is pressing on the ``heavy" jet.
	
	The growth of the RTI is seen in the two right most panels of Figure \ref{fig:hydro_Cuts}a. The panels show the condition in the jet when it is first collimated by the cocoon. The collimation shock is seen as a circular boundary that separates the unshocked jet (yellow/green) from the shocked jet material shown as a red ring. It is surrounded by the inner cocoon with a light blue-green color. RTI begins to grow on the jet-cocoon boundary just above the collimation point where the cocoon pressure compresses the jet. It shows distinctive fingers of jet material that penetrates the cocoon together with ``mushrooms" of cocoon material that develop in the jet.
	The collimation shock converges to the axis at $ z\simeq 0.1R_\star$ and reflects back onto the cocoon. The outward motion of the shock drives Richmeyer-Meshkov instabilities (RMI; \citealt{Richtmyer1960,Meshkov1969}), which accelerate the growth of the RTI fingers (Third panel from the right in Figure \ref{fig:hydro_Cuts}). The growth of the two types of instabilities can be seen in a video \href{http://www.astro.tau.ac.il/~ore/instabilities.html}{here}\footnote{\url{http://www.astro.tau.ac.il/~ore/instabilities.html}}, and was also seen in works by (e.g. \citealt{Matsumoto2013a,Matsumoto2013,Matsumoto2017,Toma2017}).
	The combination of RTI and RMI erode the jet at a faster rate and forms the JCI. Close to the jet head the baryon contamination from the cocoon becomes detrimental to the jet's integrity, and the jet becomes highly diffused with $ u_\infty \ll u_{\infty,\mx} $ (left most panel in Figure \ref{fig:hydro_Cuts}a)\footnote{Another instability that may take place along the boundary is KHI. However, relativistic KHI, which take places on the  $ \hat{x}-\hat{y} $, horizontal plane \citep{Bodo2004}, do not seem to be growing fast enough to  be present in our simulations.}.
	
	Figure (\ref{fig:hydro_Cuts}d), depicts the total energy in the box divided to logarithmic bins of $ u_\infty$ (excluding the energy of jet material that did not cross the collimation shock yet). The jet is shown in red ($ u_\infty \gtrsim 100 $), the JCI is in yellow-green ($ 3 \lesssim u_\infty \lesssim 100 $), and the blue-black colors ($ 0.1 \lesssim u_\infty \lesssim 3 $) mark the inner cocoon. The outer-cocoon maintains $ u_\infty \lesssim 0.1 $ and is outside of the $\hat{x}$ axis range.
	The intense mixing leads to a rather flat energy distribution in the logarithmic proper-velocity space (varies after breakout, see \S\ref{sec:structure}), in agreement with previous results of \citet{Gottlieb2018a}. Namely the jet does not retain most of its energy, which is roughly distributed equally on a logarithmic scale of $u_\infty$.
	
	The degree of mixing in the JCI is set by the strength of the instabilities that grow on the jet-cocoon boundary.
	In terms of the initial conditions, we find that the degree of mixing increases when the jet's opening angle is larger, its specific enthalpy is higher and its luminosity is lower. Similarly higher medium density also increases the mixing. Note that larger opening angle, lower luminosity and higher density are all leading to a slower head velocity \citep{Bromberg2011b}. The increased mixing has an additional minor effect on the head velocity\footnote{The full dependencies of the head velocity on the parameters are given by the analytic expression in \citet{Harrison2018}. Note however, that they did not explore the dependence of the head velocity on the degree of mixing. More stable jets keep their cross section at the head smaller, and thus their head velocity is larger. Here we find that the mixing can affect the velocity of the head by a factor of order unity compared to the expression of \citet{Harrison2018}.}.
	
	In sGRBs the density of the merger ejecta is significantly lower than that of stellar envelopes. This results in much stabler jets. Thus, while sGRB jets do show mixing and a significant JCI layer, typically, the central part of their core remains intact (see \S\ref{sec:sgrbs}).  
	Finally, magnetized jets may show different characteristics than the jets studied here. In the companion paper \citep{Gottlieb2020b} we show that the evolution of weakly magnetized jets is different as magnetic fields stabilize these jets.
	
	\section{The post-breakout structure}
	\label{sec:structure}

	As the jet head reaches the edge of the dense medium, it experiences a sharp drop in the density and accelerates to a velocity close to the speed of light. After breakout, both the jet and the cocoon continue to accelerate under their own pressure and expand sideways where each component moving at $ \Gamma\beta $ expands to an opening angle $ \theta \sim \rm{atan}(\frac{1}{\Gamma\beta}) $.
	In what follows we model the angular distribution of the jet-cocoon system.
	We consider both the temporal evolution and the final distribution from which one can infer characteristics of the prompt and afterglow emissions, respectively.
	
	We generally find that the structure of the outflow is composed of three regions that correspond to the structure of the jet-cocoon system before breakout (see bottom panel in Figure \ref{fig:2d3dmaps} for a visual illustration):\\
	(i) The jet core: characterized by ultra-relativistic velocities, $ u_\infty \gtrsim \frac{1}{5}u_{\infty,\mx} $, and an isotropic equivalent energy distribution, $ \Eiso $ with a rather flat angular profile. We define the core angle $\theta_j$ as the angle in which $ \Eiso $ drops to 75\% of its value on the axis.  We find that the jet core becomes slightly narrower over time and stabilizes at $ \theta_j \approx (\frac{1}{3} - \frac{1}{5})\theta_{j,0} $\footnote{The higher values of $ \theta_j/\theta_{j,0} $ are obtained in less massive media, such as in sGRBs.}, in agreement with the result of \citet{Mizuta2013}. \\
	(ii) The cocoon: material with Newtonian to mildly-relativistic velocities, $ u_\infty \lesssim 3 $ expanding at large angles, $ \theta \gtrsim \theta_c $, where $ \theta_c \approx 0.3 \rad $, as expected for $ u_\infty \lesssim 3 $.
	The cocoon receives its energy during the jet propagation inside the dense medium and its total energy $\sim L_jt_b$, where $t_b$ is the breakout time\footnote{A more accurate approximation to the cocoon energy is $E_c \sim L_j(t_b-R_*/c)$, accounting for the energy that remains in the unshocked jet at the time of the breakout. 
		However, in most GRBs (and our simulations), the head is expected to be subrelativistic so $R_*/c \ll t_b$ and it can be neglected.}. \\
	(iii) The jet-cocoon interface (JCI): this part is composed of the mixed material resulting from the jet-cocoon interaction that takes place inside of the dense medium. Since the cocoon pressure drops slowly with time, the jet collimation and the associated mixing in the JCI continues also after the breakout of the jet from the dense medium. In our models the mixing is relatively intense and the JCI contains a considerable amount of the outflow energy. Hence, it can play an important role in the hydrodynamical evolution and the resulting emission. 
	Typically the JCI stretches from mildly-relativistic velocities to ultra relativistic velocities, and lies at $ \theta_j \lesssim \theta \lesssim \theta_c $.
	
	In the following discussion of the distributions we address each of the components separately. To ease the differentiation between them, the background color of the distribution plots is painted in pale blue for the jet region, pale red for the JCI, and pale yellow for the cocoon.
	
	To a good approximation all of the jet energy prior to the breakout goes into the cocoon, while the energy deposited after the breakout is divided between the jet core and the JCI. Thus, defining the time after the breakout as $T \equiv t-t_b$, the total energy in the jet and in the JCI at any given time is $E_j + E_{JCI} \sim L_j T$. The fraction of the energy that goes to the JCI is mainly determined by the mixing. We therefore define the mixing parameter to be the fraction of jet core energy from the total injected energy after breakout, $ \lambda \equiv E_j/L_j T$.
	In our simulations we find that the angular distributions of the isotropic equivalent energy of the JCI can be approximated by a power-law $ \Eiso \propto \theta^{-\delta} $. By integrating the total energy, one obtains that for $ \delta > 2 $, $ \delta \approx \frac{2}{1-\lambda} $. For $ \delta < 2 $ the relation between $ \lambda $ and $ \delta $ can be obtained numerically.
	While the value of $ \lambda $ changes with time, we find that typically lGRB jets (\S \ref{sec:lgrbs}), which undergo larger mixing, maintain values of $ \lambda \sim 0.1-0.2 $ with all models featuring $ \delta \lesssim 2 $.
	Short GRB jets (\S \ref{sec:sgrbs}), on the other hand, can be more stable with a core energy that is comparable to the energy in the JCI so that $ \lambda \approx 0.4 $, and hence $ \delta \sim 3 $. The values of $ \lambda $ and $ \delta $ of the different models are listed in Table \ref{tab_summary}. We also consider one model of a lower luminosity sGRB, which features an evolution that is similar to that of long GRBs as we discuss in \S \ref{sec:low_luminosity_sgrbs}.
	\subsection{Long GRBs}
	\label{sec:lgrbs}
	
	Long GRB jets, propagating in massive stars, typically have slow head velocities prior to the breakout. Consequently the degree of mixing in these systems is generally higher than that of short GRBs, which propagate faster). After the breakout, as time progresses, more and more stellar material is evacuated from the medium surrounding the jet so the pressure applied on the jet drops and so does the mixing experienced by freshly launched jet material. Under extreme conditions, e.g. highly narrow and powerful jets, an inverse evolution in time is observed. Namely, the mixing is low prior to breakout and it increases with time instead of decreasing. An example to such a system is model $ \F $, which is briefly discussed in this paper.
	Weakly magnetized jets display a similarly small mixing while propagating inside stars. The conditions that lead to such a behavior are shared by weakly magnetized jets and extremely powerful and narrow hydrodynamic jets, and are discussed in a companion paper\footnote{Note that unlike hydrodynamic jets, weakly magnetized jets continue to maintain low level mixing also after breakout.} \citep{Gottlieb2020b}. Here we focus on characterizing hydrodynamic lGRB jets with common parameters.
	
	\subsubsection{Energy distribution in velocity space}
	
	\begin{figure}
		\centering
		\includegraphics[scale=0.225]{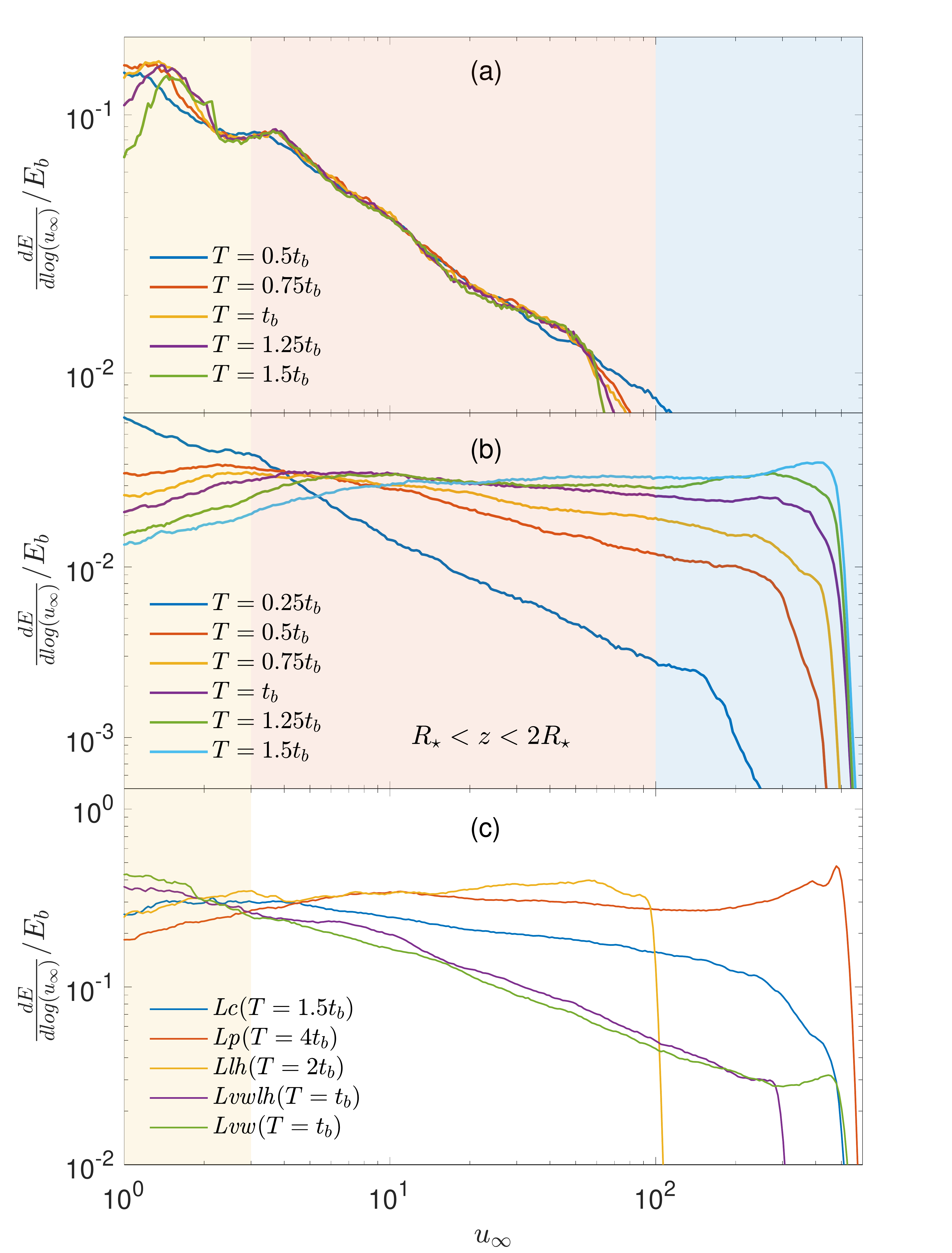}
		\caption[$ E(u_\infty) $]{
			The energy distribution per logarithm of the terminal proper-velocity, $ u_\infty = h\Gamma - 1 $ of matter that broke out from the star.
			(a) \textbf{The temporal evolution of a given slab of matter (Lagrangian) at different times.}. The distribution of the first slab of matter that broke out form the stellar envelope in simulation $ \A $. The slab length is $R_\star$.
			The amount of mixing does not change at these times and the distribution remains unchanged, except for some evolution at low and high $ u_\infty$ (see text).		
			(b) \textbf{The change in the mixing between different slabs of matter.} We present the slabs that are located at $ R_\star < z < 2R_\star $ in simulation $ \A $ at different times. Fresh elements undergo less mixing and are able to maintain higher $ u_\infty$.
			(c) \textbf{The mixing in different simulations of the matter}. The distributions include all the material that broke from the star when the front of the jet reaches $ 10R_\star $. Note that the total mixing in model $ \A $ is not as flat as the matter that broke out last in (b), due to the contribution of more contaminated matter in the jet front.
			The distributions are normalized by the energy upon breakout $ E_b $.
		}
		\label{fig:Ehg}
	\end{figure}
	
	{In the previous section we discussed 
		the jet-cocoon energy distribution as a function of $ u_\infty$ while the jet head propagates in the star (Figure \ref{fig:hydro_Cuts}d depicts
		this distribution in simulation $ \A $). Here we discuss the same distribution for lGRBs at various times after the jet breakout. 
		Figure \ref{fig:Ehg} depicts the temporal evolution of the energy distribution  for matter that broke out from the star.
		Figure \ref{fig:Ehg}a traces the distribution in the first slab of matter that breaks of the star at different times in simulation $ \A $, showing that the energy distribution in $ u_\infty $ remains essentially the same, particularly in the JCI. 
		The lack of change in the mixing in the slab indicates that $ u_\infty $ at the breakout is a good approximation of the true terminal velocity of a fluid element.
		The energy distribution at low and high $ u_\infty $ does show some evolution with time after the breakout.
		The evolution at the low end, ($ u_\infty \lesssim 3 $) is an artifact of the simplified method we use to trace in time a Lagrangian relativistic slab with an Eulerian code. We assume it moves at $ v = c $ and therefore this method cannot trace material with mildly-relativistic velocities.
		At the high end of $u_\infty$ the differences are physical and originate in internal shocks between jet elements with different velocities. Consequently, elements with low $ u_\infty $ are shocked by faster elements, so that energy is shifted from the high $ u_\infty $ tail to slower velocities.
		
		When looking at the evolution of the system by observing different slabs that go through the region $ R_\star < z < 2R_\star $ in simulation $ \A $ (Figure \ref{fig:Ehg}b), substantial differences are found among elements.
		Most of the cocoon energy originates in  matter that breaks out over a duration $T \sim t_b $, after which the energy distribution in the cocoon is roughly constant, with a total energy $ \sim L_jt_b $ that does not change by more than a factor of 1.5. 
		The energy distribution at $ u_\infty\gtrsim 3 $, is part of the jet and the JCI. It has a power-law shape with a cutoff that evolves to become flatter with time.  The first slab suffers the highest mixing, as it is affected also by the high mixing at the jet head.
		At later times the pressure in the cocoon drops due to the depletion of cocoon energy, and the jet slowly becomes conical. The jet-cocoon interaction weakens towards a new steady state where the mixing at the JCI maintains constant energy per logarithmic velocity interval at the JCI and the jet.
		
		The energy distribution per logarithm of $u_\infty$ in most lGRB models show a qualitatively similar temporal evolution.
		The major difference between the models is the initial power-law distribution with which the matter breaks out from the star. It is steeper if the mixing is high, and plateaus faster if the mixing is low.
		Eventually all models reach a roughly flat distribution with a cut-off at $ \sim u_{\infty,\mx} $.
		In Figure \ref{fig:Ehg}c we present the energy distribution in a sample of models when the jet head reaches $ 10R_\star $. The energy distributions in models $ \A, \I $ and $ \J $ are not flat since $ T \approx t_b $, so that only the highly mixed material broke out and thus it has less energy at high $u_\infty$, as seen for example in Figure \ref{fig:Ehg}a. However, the slabs in the rear of these jets in all the simulations already show a flat distribution, as seen for model $ \A $ in Figure \ref{fig:Ehg}b.
		We find that model $ \F $ shows a different evolution from other models. It includes a very narrow and extremely energetic jet with $ \Liso = L_j\theta_{j,0}^{-2} \approx 1.5 \times 10^{53} \erg\s^{-1} $ along the jet axis, which is seen only rarely in lGRBs. The jet in this simulation is relatively stable at first with a distinct peak of the energy distribution at $ u_\infty = u_{\infty,\mx} $. At late times however, the mixing inside the star strengthens rather than diminishes during the expansion of the collimation shock. Consequently, similar to all other models, its terminal energy distribution is rather flat. The origin of this behaviour is a highly pressurized structure that forms near the base of the jet and affects the collimation shock. It is formed soon after the jet launching starts and it is dissolved as the pressure in the cocoon starts to drop. This structure (and this evolution) is not seen in any of the other simulations, where the conditions are more similar to those seen in typical lGRBs. It is rather common, though, in weakly magnetized jets and we discuss it in detail in \citet{Gottlieb2020b}.
		
		To conclude, the energy distribution with $u_\infty$ can be modeled with two components: the cocoon at $ u_\infty \lesssim 3 $ and a power-law at $ u_\infty \gtrsim 3 $ that evolves with time until a plateau is reached.
		Models with lower (higher) mixing show smaller (larger) initial power-law indices. By $ T \approx 2t_b $ all models reach a quasi-flat distribution with a cut-off at $ \sim u_{\infty,\mx} $.
		At $ T \gtrsim t_b $ the total energy in the cocoon is roughly constant, $ E_t(u_\infty \lesssim 3) \approx L_jt_b $, while the total energy in the jet and the JCI is $ E_t(u_\infty \gtrsim 3) \approx L_jT = L_j(t-t_b)$.
		
		\subsubsection{Angular Distribution of the Energy}
		
		\begin{figure}
			\centering
			\includegraphics[scale=0.23]{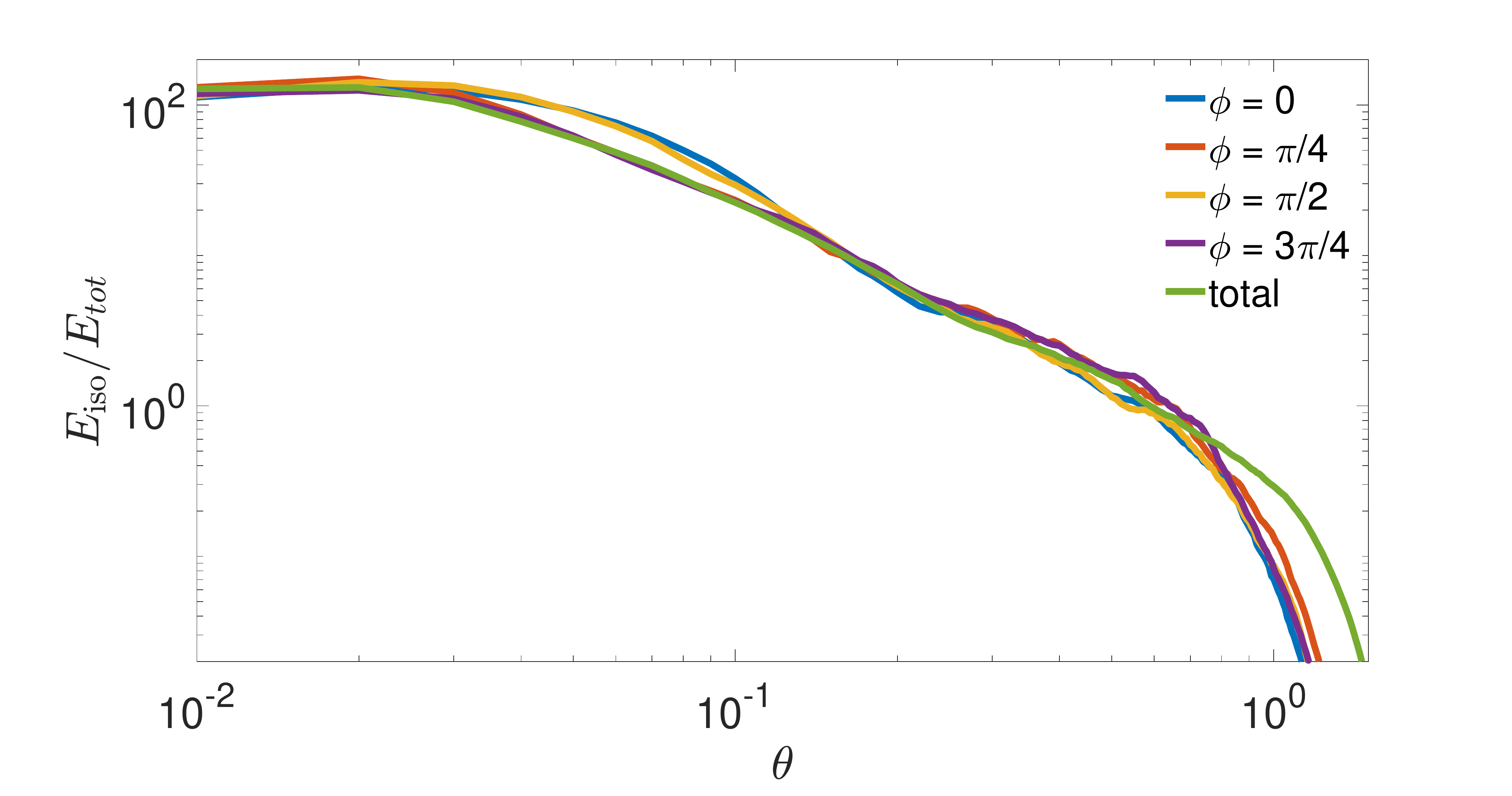}
			\caption[$ \Eiso(\theta) $]{
				A comparison of the total angular distribution of model $ \A $ with 2D plane distributions. Taken when the jet reaches $ 10R_\star $. }
			\label{fig:angular_cuts_comparison}
		\end{figure}
		
		\begin{figure}
			\centering
			\includegraphics[scale=0.23]{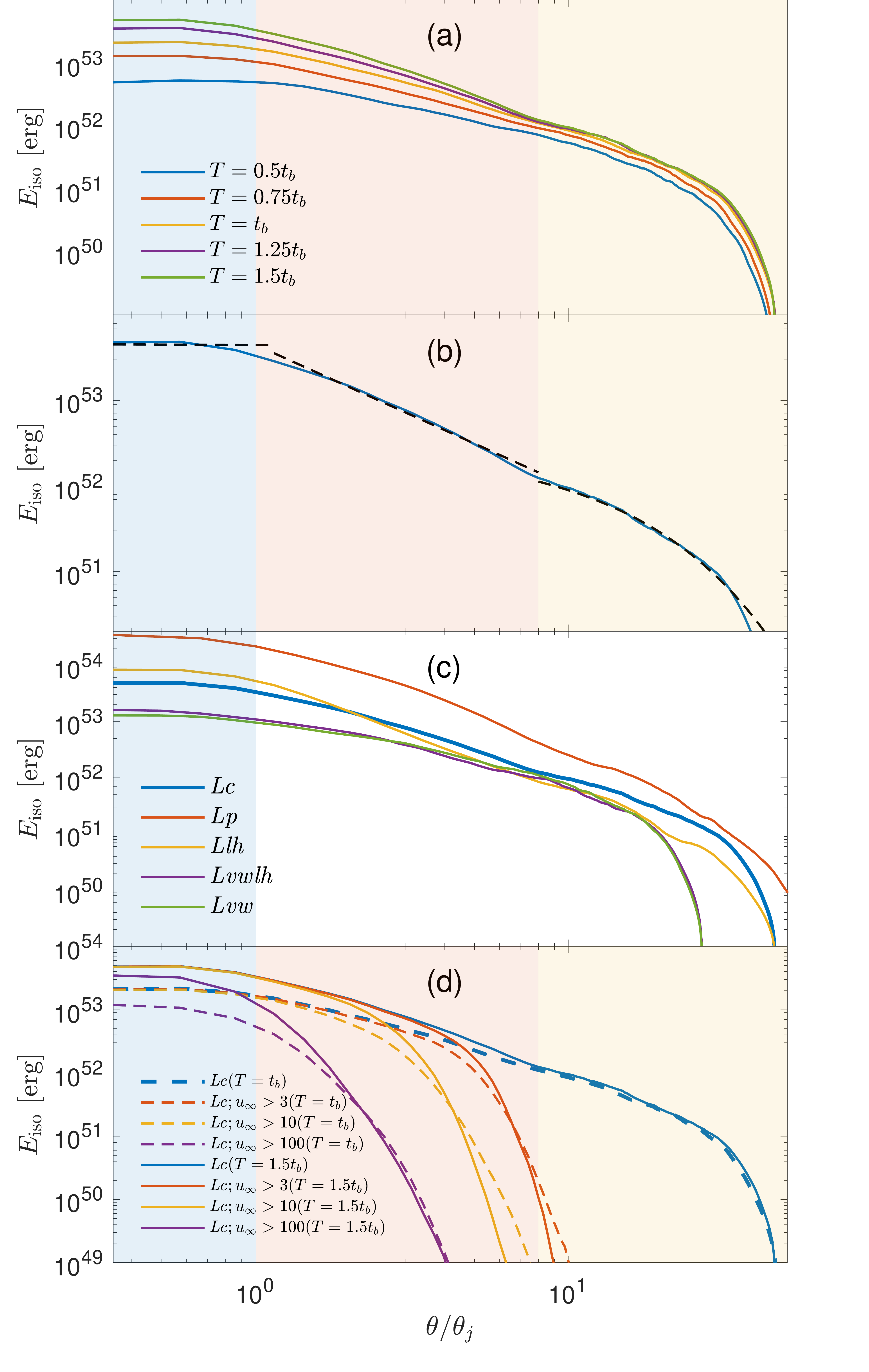}
			\includegraphics[scale=0.23]{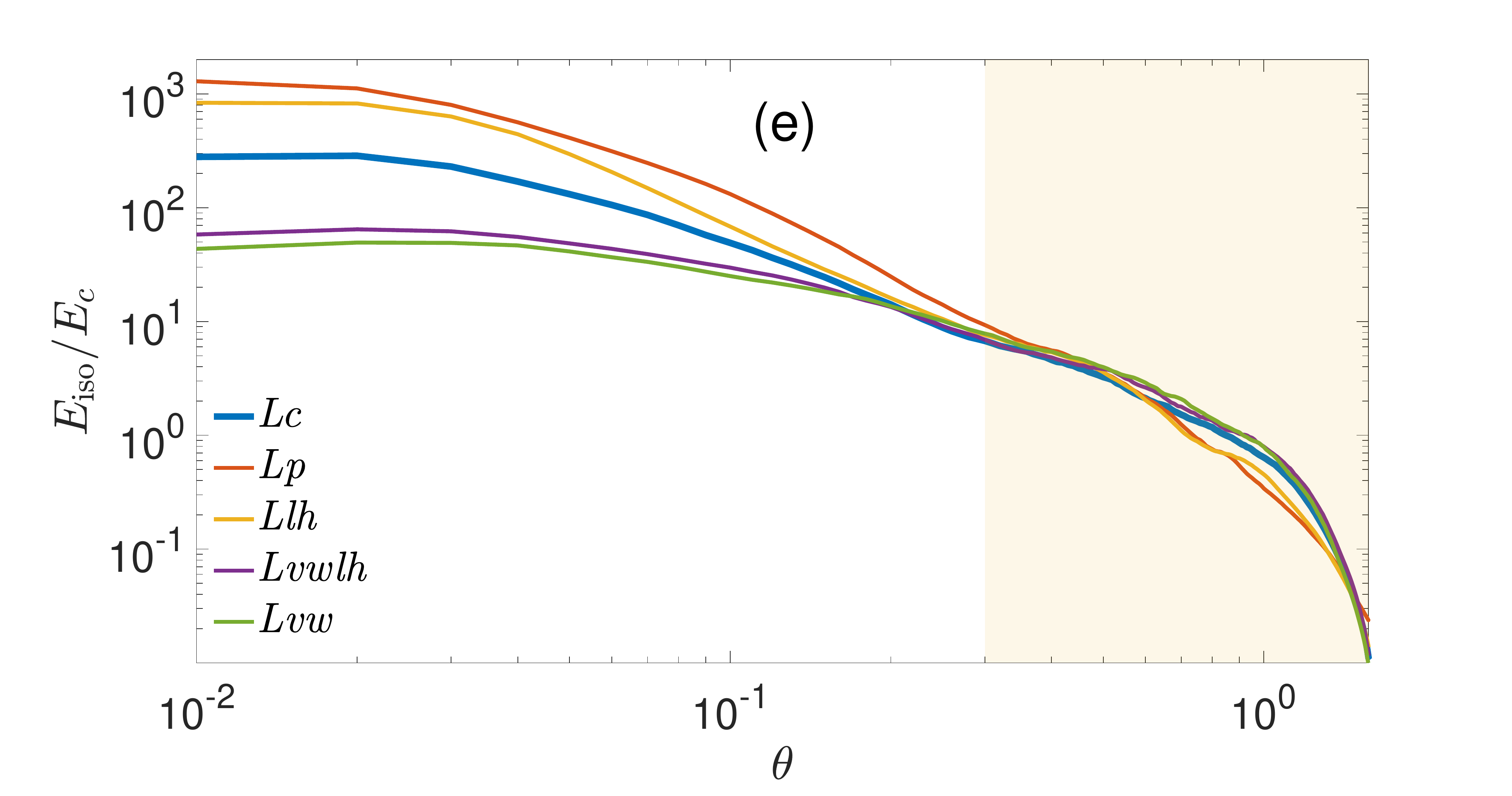}
			\caption[$ \Eiso(\theta) $]{
				The angular distribution of the isotropic equivalent energy in lGRBs for matter that broke out from the star.
				(a) A comparison between distributions of model $\A $ at different times.
				(b) Analytic fits (dashed black lines) to the three components: jet, jet-cocoon interface and cocoon in model $ \A $.
				(c) A comparison between different models when the jet reaches $ 10R_\star $
				(d) Distributions of matter faster than a designated 4-velocity in model $ \A $.
				(e) Similar to (c), but the energy is normalized by the cocoon energy, and the distribution is given as a function of $ \theta $ to manifest the similarity of the cocoons in the different models.
			}
			\label{fig:Etheta}
			
		\end{figure}
		
		We inject an axisysemtric jet into an axisymetric medium. Yet, the jet that emerges from the medium is non-axisymmetric due to the stochastic loading which is induced by the instabilities. We test first how strong the symmetry breaking is. Figure \ref{fig:angular_cuts_comparison} shows the distribution of the isotropic equivalent energy ($4\pi\times dE/d\Omega$) as a function of the polar angle $\theta$ along 2D cuts at four different azimuthal angles. These cuts are compared to the distribution obtained by averaging over the azimuthal angle. It shows that there are differences between the various cuts but they are rather minor. Thus, the axisymmetry breaking is not severe and the outflow can be approximated as being axisymmetric, at least for the purpose of the distributions discussed here. In the following all the presented distributions are the averages over the azimuthal angle.
		
		Figure \ref{fig:Etheta} depicts the angular distribution of the isotropic equivalent energy.
		Figure \ref{fig:Etheta}a shows the temporal evolution of the distribution in the canonical model $ \A $. It highlights the three components of the system by the background colors (a fit is provided in Figure \ref{fig:Etheta}b):
		(i) A flat jet core ($ \theta < \theta_j $). Its energy grows with time as $ \lambda L_jT $, where $\lambda$ is the mixing parameter, as more jet material is injected into the system. Note however that since the degree of mixing changes with time, the fraction of energy deposited in the core is not necessarily constant and hence $ \lambda L_jT $ does not have to grow linearly with time.
		(ii) The cocoon  ($ \theta>\theta_c = 0.3\rad $ ; Figure \ref{fig:Etheta}e). As the cocoon includes also non-relativistic material, most of its material breaks out continuously until $ T \sim t_b $. At later times its energy changes only by a factor of order unity as the slowest material ($u_\infty \sim 0.1$) from the outer cocoon (i.e., shocked medium) continues to emerge from the star. 
		We find that in all models the segment of the cocoon's energy can be well described by an exponential decay, $ \Eiso \propto e^{-f_c\theta} $, where $ f_c $ is reduced with the cocoon energy. For our lGRB models $ 1 \lesssim f_c \lesssim 3 $ (the values of $ f_c $ are given in Table \ref{tab_summary}).
		(iii) The transition from the jet to the cocoon at the jet-cocoon interface (JCI) which lies at $ \theta_j < \theta \lesssim \theta_c $. The energy distribution in this segment can be approximated by a power-law distribution, with a power-law index $ -\delta $. The value of $ \delta $ increases with time as a consequence of the decrease in the mixing at the jet boundary. In general heavy baryon loading leads to smaller $\lambda$ and smaller $\delta$, namely it flattens the distribution, as shown in Figure \ref{fig:Etheta}c. Note that while the jet core and the JCI are characterized better as a function of $ \theta/\theta_{j,0} $, the cocoon is described better as a function of $\theta$. Thus in Figure \ref{fig:Etheta} panels (a-d) are shown in $ \theta/\theta_{j,0} $ and panel (e) is given in radians, showing the similarity between the cocoons in different models.
		
		In conclusion, at $ T \gtrsim t_b $ the energy in the cocoon remains fixed, and all the injected energy is divided between the jet and the JCI. The distribution takes the form (see fit in Figure \ref{fig:Etheta}b)
		\begin{equation}\label{eq:Etheta}
		\Eiso(\lambda) \approx
		\begin{cases}
		\lambda E_0       & { \theta < \theta_j }\\
		\lambda E_0 (\theta/\theta_j)^{-\delta}  & { \theta_j < \theta < \theta_c }\\
		\lambda E_0 (\theta_c/\theta_j)^{-\delta}e^{-f_c(\theta-\theta_c)} & { \theta > \theta_c } \ ,
		\end{cases}
		\end{equation}
		where $ E_0 \equiv \frac{L_jT}{1-{\rm cos}\theta_j}$.
		Once the jet engine shuts off, the distribution does not change substantially. The jet core no longer evolves as no fresh, relativistic jet material breaks out. However, as more mildly-relativistic material breaks out from the star, the other parts do show some evolution, although not a very significant one.
		
		Finally, during the afterglow phase the relevant energy is the one deposited in the fastest material as the outflow decelerates. For that reason, we plot the angular distribution of isotropic equivalent energies at velocities larger than a given value of $ u_\infty $ (Figure \ref{fig:Etheta}d). We plot in dashed line the distributions of model $ \A $ at $t=t_b$ and in solid lines at $t=1.5t_b$.
		We find that for each value of velocity, $ u_i $, there is a corresponding angle, $ \theta_i $, below which the matter with $ u_\infty > u_i $ dominates, and above it matter with $ u_\infty > u_i $ is negligible. At $\theta<\theta_i$ the energy distribution follows the same distribution of the total energy in Equation \ref{eq:Etheta} and it cuts off at $\theta>\theta_i$\footnote{Over time, the drop steepens as can be seen in the figure. By $ t \gg t_e $ which is the relevant time for the afterglow, the drop is essentially a step function.}.
		The angle at which the decline of $ u_i $ takes place is the angle that corresponds to $ u_i $ in Figure \ref{fig:vtheta}. For example, for the cocoon which begins at $ \theta_c = 0.3\rad \approx 8\theta_j $, Figure \ref{fig:vtheta} shows that $ <u_\infty> \approx 3$, this is also where the red line in Figure \ref{fig:Etheta}d drops.
		
		\subsubsection{Angular Distribution of the Velocity}
		
		\begin{figure}
			\centering
			\includegraphics[scale=0.23]{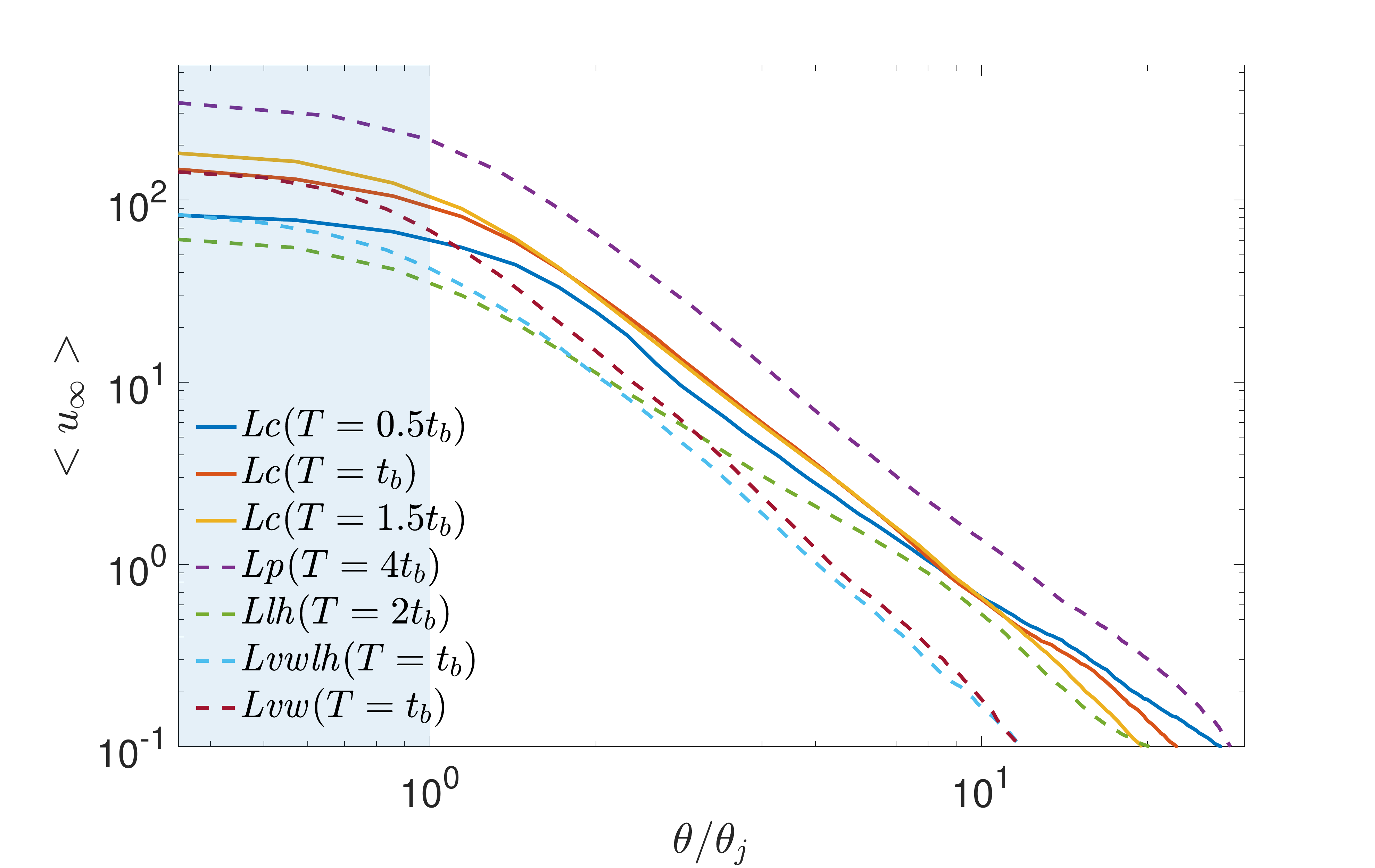}
			\caption[$ u_\infty(\theta) $]{
				The angular distribution of the energy-weighted average of the terminal proper-velocity, $ <u_\infty>(\theta) \equiv \int_\theta{u_\infty dE}/\int_\theta{dE}$, given for different models and at different times. We consider only matter that broke out from the star.
				The continuous lines reflect the temporal evolution of the distribution of model $ \A $. The dashed lines show different models.
			}
			\label{fig:vtheta}
		\end{figure}
		
		\begin{figure}
			\centering
			\includegraphics[scale=0.23]{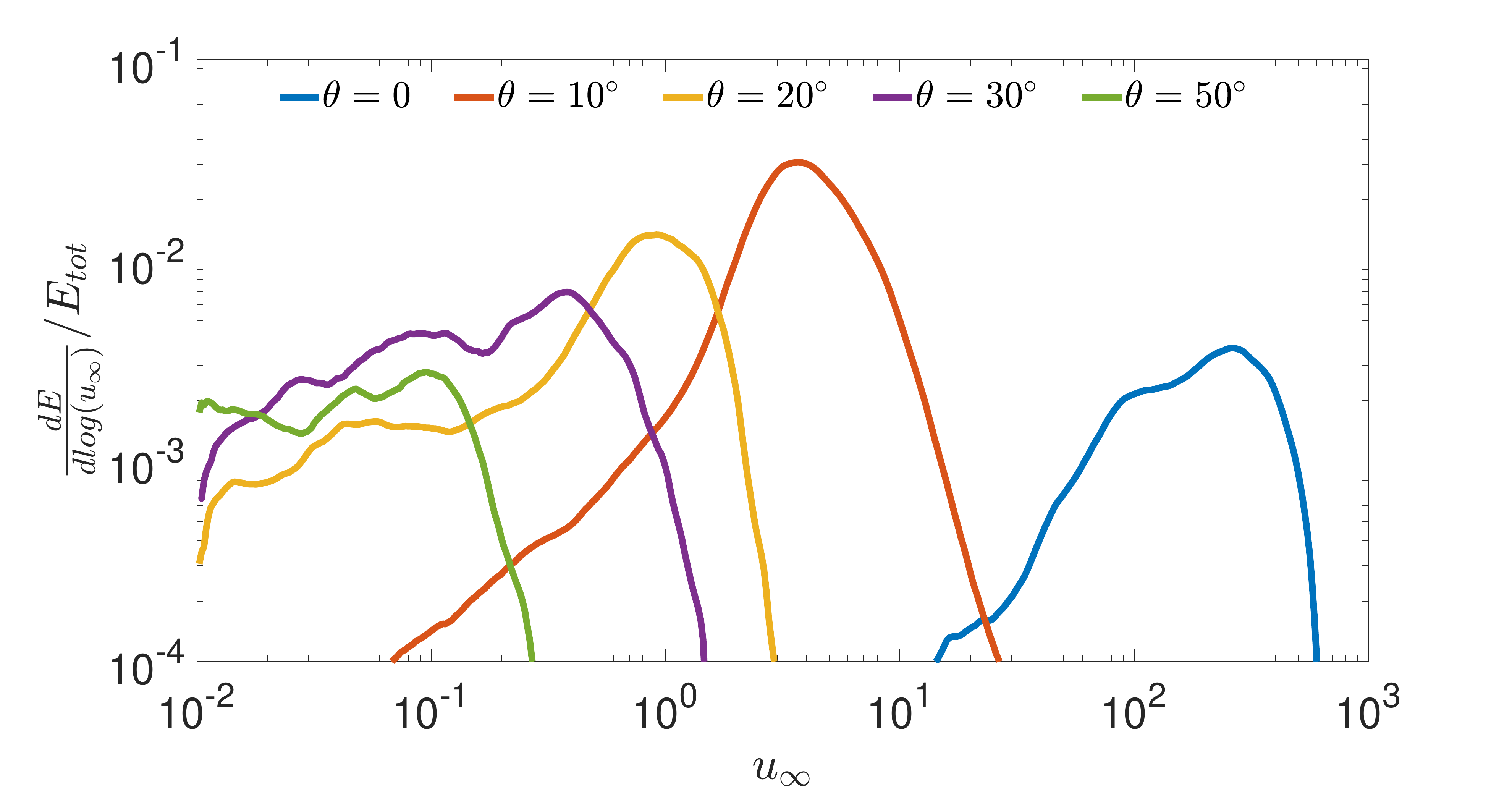}
			\caption[$ E(u_\infty) $]{
				The energy distribution per logarithmic scale of $u_\infty$ at certain angles, normalized by the total energy at the given angle.
				The simulation presented is $ \A $ and each distribution includes only matter that broke out from the star at $ T = 1.5t_b $. The distinct peaks at different angles of the jet and the JCI demonstrates that the distribution is dominated by the angular structure rather than a radial one in the component. The distribution in the cocoon ($\theta>20^\circ$) is almost constant in logarithmic space, implying that the cocoon has also a radial structure.
			}
			\label{fig:Eu_angle}
		\end{figure}
		
		A general axisymmetric homologous outflow has a two dimensional structure, angular and radial, and it should be described by a 2D density profile, $\rho(\theta,u)$. However, the structure of GRB jets is often approximated as having only an angular structure. Namely, at any direction the outflow has a characteristic proper-velocity and all the mass is concentrated in a thin shell so there is no radial structure. In that case the outflow velocity can be described by a one dimensional profile $u(\theta)$. Figure \ref{fig:Eu_angle} shows the energy distribution per logarithmic scale of $u_\infty$ along specific directions. It shows that at the core and the JCI ($\theta \lesssim 20^\circ$), there is a characteristic velocity along each direction and therefore these components have no significant radial structure. The cocoon ($\theta > 20^\circ$), however,  has a rather flat distribution of energy in $u_\infty$ log space, and therefore it has both a radial and an angular structure. Below, we focus on $u_\infty(\theta)$ of the core and the JCI.
		
		Figure \ref{fig:vtheta} shows the angular distribution of the energy-weighted average of the terminal proper-velocity, $ <u_\infty> $.
		All models share the same behavior with a roughly flat core at $ \theta \lesssim \theta_j $, followed by a power-law distribution:
		\begin{equation}\label{eq:utheta}
		u_\infty(\theta/\theta_j) =
		\begin{cases}
		<u_{\infty,j}>   & { \theta < \theta_j }\\
		<u_{\infty,j}> (\theta/\theta_j)^{-p_u}   & { \theta > \theta_j } \ .
		\end{cases}
		\end{equation}
		The jet breaks out from the star with an average four velocity at its core of $ <u_{\infty,j}> $. The value of
		$ <u_{\infty,j}> $ depends on the mixing, with lower mixing yields higher $ <u_{\infty,j}>/u_{\infty,\mx} $.
		In our models we find that $ <u_{\infty,j}> \approx 2\lambda u_{\infty,\mx} $.
		Outside the core the power-law index increases with time as the mixing drops and the terminal Lorentz factor at the core increases. For example, upon breakout model $ \A $ features $ p_u \approx 1.6 $, which increases to $ p_u \approx 2 $ when the head reaches $R_*/4$ and to $ p_u \approx 2.4 $ when the head reaches $R_*/10$. The values of $p_u$ for all the models are listed in Table \ref{tab_summary}.
		
		\begin{table}
			\setlength{\tabcolsep}{6pt}
			\centering
			\begin{tabular}{ | l | c  c  c  c | }
				
				\hline
				lGRB Model & $ \lambda_4~(\lambda_{10}) $ & $ \delta_4~(\delta_{10}) $  & $ f_{c,4}~(f_{c,10}) $ & $ p_{u,4}~(p_{u,10}) $\\ \hline
				$ \A $ & 0.1 (0.13) & 1.2 (1.8) & 1.6 (1.5) & 2.0 (2.4) \\
				$ \B $ & 0.09 & 0.8 & 1.3 & 2.1 \\
				$ \C $ & 0.18 & 1.8 & 2.6 & 1.8 \\
				$ \D $ & 0.15 (0.19) & 1.7 (2.2) & 2.8 (2.1) & 1.9 (2.3) \\
				$ \E $ & 0.16 & 1.7 & 2.5 & 2.1 \\
				$ \F $ & 0.27 (0.14) & 2.2 (1.9) & 2.5 (2.5) & 2.3 (2.5) \\
				$ \G $ & 0.19 & 1.2 & 1.7 & 2.0 \\
				$ \h $ & 0.13 (0.25) & 1.4 (2.0) & 1.6 (1.9) & 1.9 (1.9) \\
				$ \I $ & 0.08 (0.11) & 0.7 (1.3) & 0.9 (1.4) & 2.1 (2.6) \\
				$ \J $ & 0.06 (0.11) & 0.7 (1.1) & 1.0 (1.4) & 2.1 (2.7) \\
				\hline
				sGRB Model & $ \lambda $ & $ \delta $  & $ f_c $ & $ p_u $\\ \hline
				$ \SG_1 $ & 0.4 & 3.1 & 4.5 & 2.7 \\
				$ \SG_2 $ & 0.43 & 3.5 & 3.7 & 2.4 \\
				$ \SG_3 $ & 0.38 & 3.2 & 4.1 & 3.2 \\\hline
				
			\end{tabular}
			\hfill\break
			
			\caption{
				A summary of the models characteristics: $ \lambda $ is the jet core energy to the total energy ratio. $ \delta $ and $ f_c $ are the indices in the isotropic equivalent energy distribution, where $ \Eiso \propto \theta^{-\delta} $ in the jet-cocoon interface, and $ \Eiso \propto e^{-f_c\theta} $ in the cocoon. $ p_u $ is the power-law index in the angular distribution of the energy-weighted average of the proper-velocity, $ < u_\infty > \propto (\theta/\theta_j)^{-p_u} $. Subscripts 4 and 10 reflect the location of the jet head, at $ 4R_\star $ and $ 10R_\star $, respectively.
				In models $ \B, \C, \E $ and $ \G $ the simulations were terminated when the jet head reached $ 4R_\star $.
			}
			\label{tab_summary}
		\end{table}

		\subsection{Short GRBs}
		\label{sec:sgrbs}
		
		\begin{figure}
			\centering
			\includegraphics[scale=0.21]{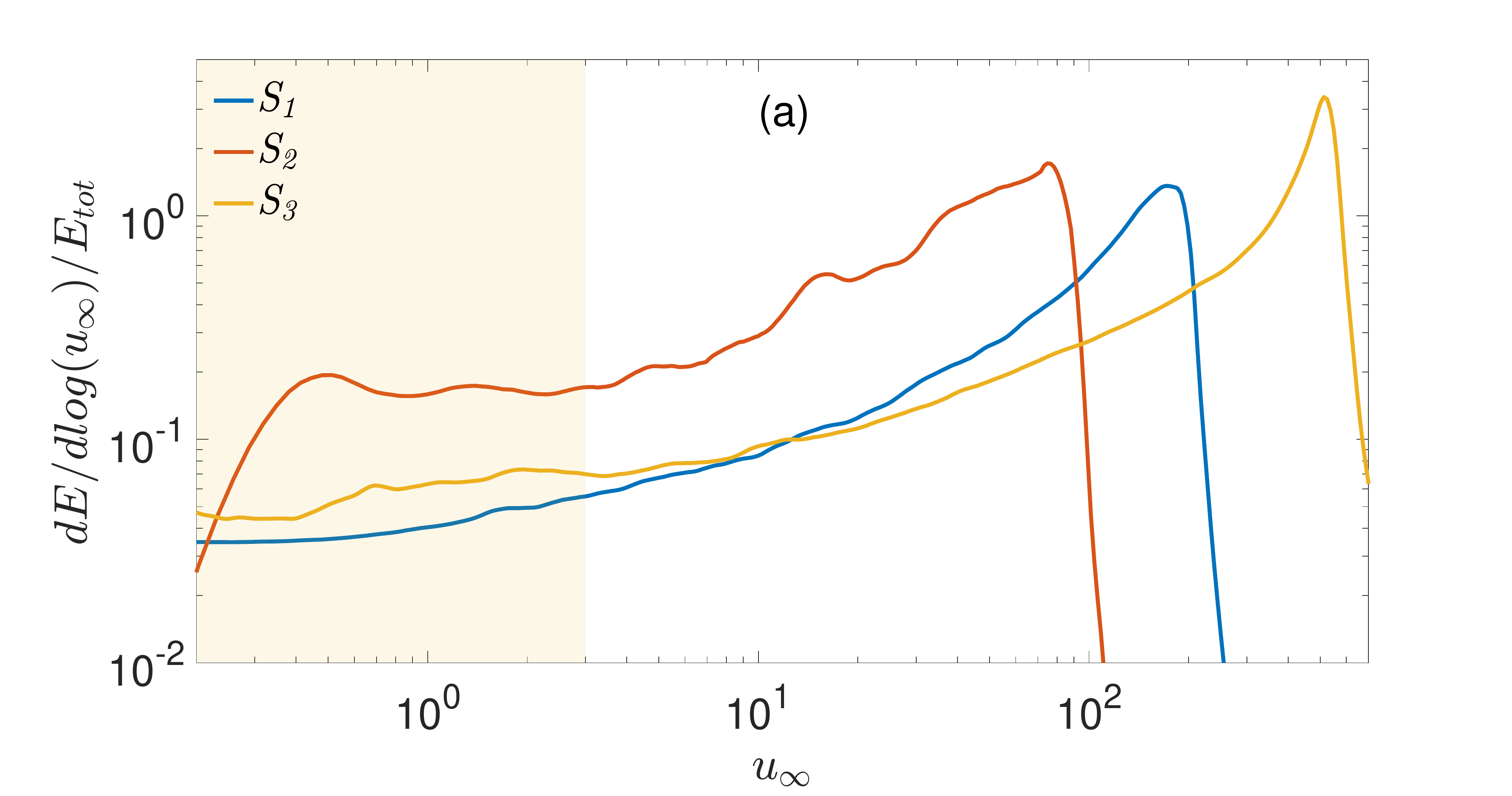}
			\includegraphics[scale=0.21]{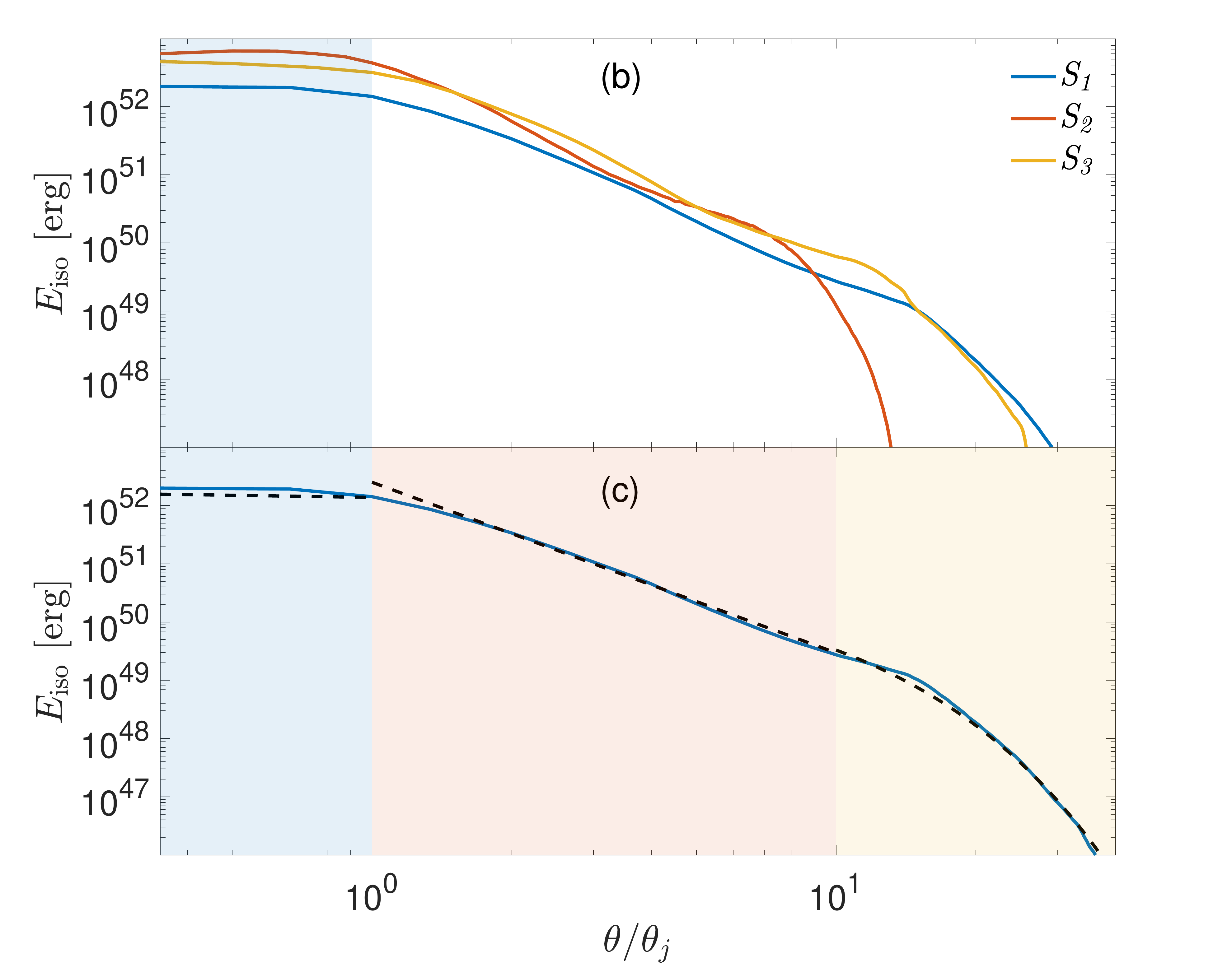}	\includegraphics[scale=0.21]{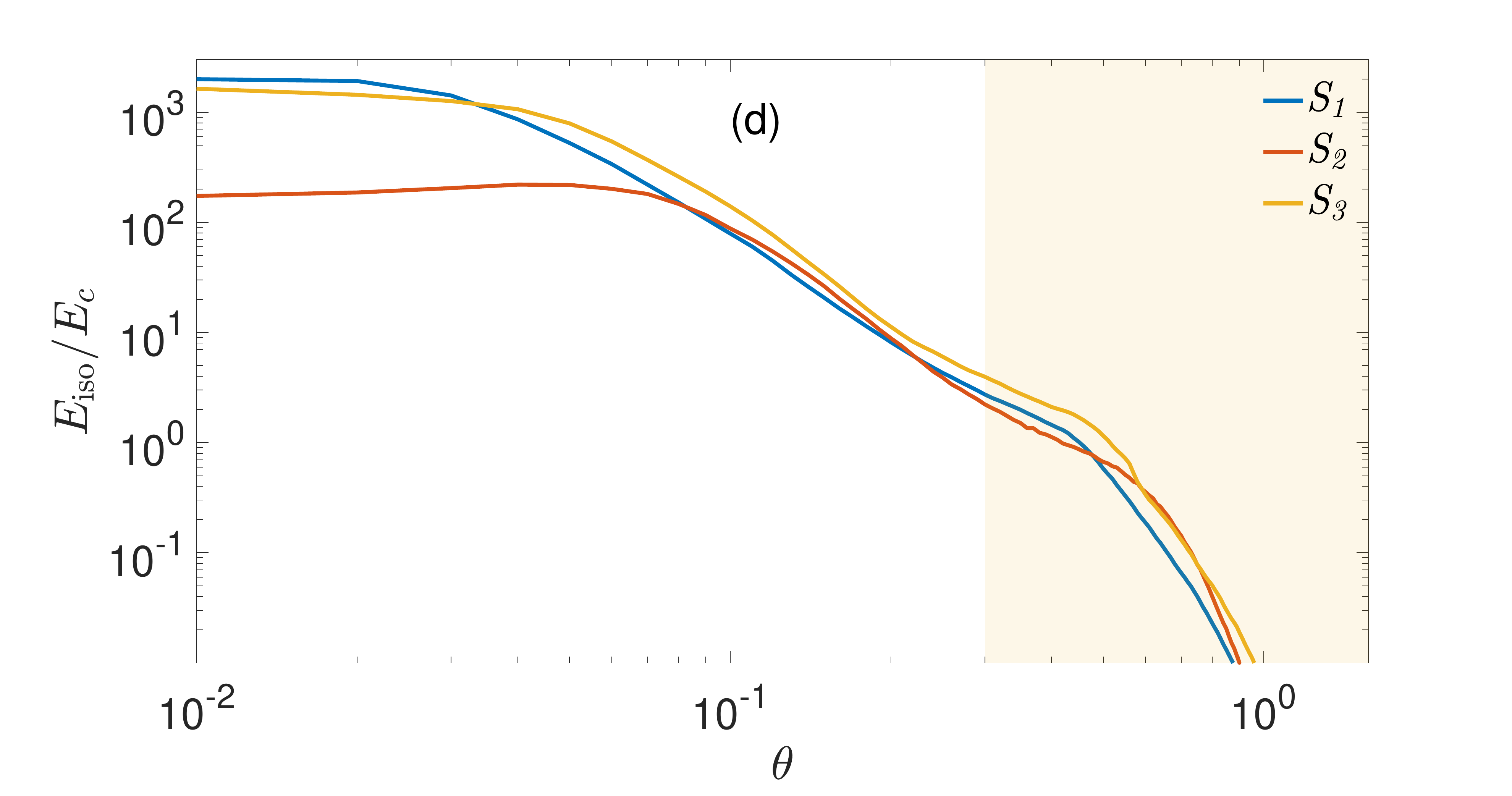}
			\includegraphics[scale=0.21]{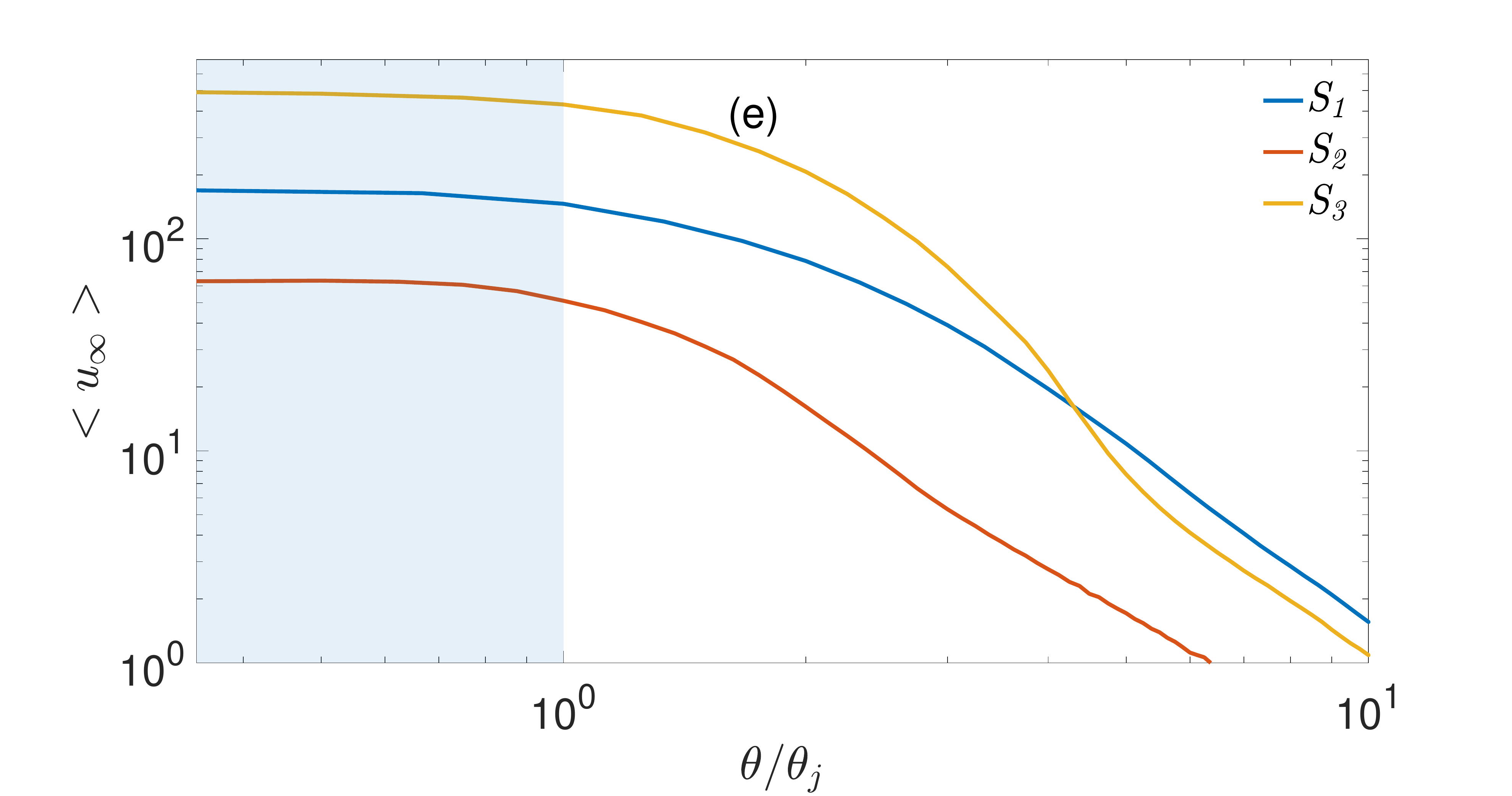}
			\caption[sGRB distributions]{
				The sGRB models' distributions at the homologous phase, for matter that broke out from the core ejecta and is above the collimation shock.
				(a) The energy distribution per logarithmic scale of proper-velocity, normalized by the total energy of each model.
				The angular distribution of the isotropic equivalent energy for all models (b) and the best fit to model $ \SG_1 $ (c).
				(d) Similar to the upper second panel, but energy normalized by the cocoon energy and given as a function of angle to demonstrate the similarity in the cocoon between models.
				(e) The angular distribution of the energy-weighted average of the terminal proper-velocity.
			}
			\label{fig:sGRB_dists}
		\end{figure}
		
		In this part we repeat the aforementioned lGRB analysis for sGRBs.
		Generally sGRBs can be modeled with similar expressions. 
		However, the distributions of models $ \SG_1, \SG_2 $ and $ \SG_3 $ show substantial differences in the values of $ \lambda $ and $\delta$ between the two GRB types.
		The differences originate from the less dense media that surrounds sGRB jets, which result in a more stable jet boundary and lower baryon contamination from the surroundings (see \S\ref{sec:evolution}). This leads to cleaner jets, which feature higher values of $ \lambda $ and $\delta$.
		Another difference between the two types is the expansion of the collimation shock out of the ejecta, which is rarely the case in lGRBs and is quite common in sGRBs. This can have important effects on the prompt emission \citep{Gottlieb2019b}.
		
		Figure \ref{fig:sGRB_dists}a shows the energy distribution per logarithmic proper-velocity.
		The cocoon and the jet-cocoon interface segments are not very different from lGRB models, possessing a mild power-law distribution in these regions.
		However, unlike lGRBs, the energy in the more stable sGRB jets increases at $ u_\infty \gtrsim \frac{1}{5}u_{\infty,\mx} $ until it peaks at $ u_\infty = u_{\infty,\mx} $.
		
		The angular isotropic equivalent energy distribution in Figure \ref{fig:sGRB_dists}b is in excellent agreement with Equation \ref{eq:Etheta} (a fit is shown in Figure \ref{fig:sGRB_dists}c).
		Two quantitative differences are seen when comparing the isotropic equivalent energy distributions of lGRBs and sGRBs. First, the power-law of the jet-cocoon interface is much steeper with $ \lambda \approx 0.4 $ and consequently $ \delta \gtrsim 3 $, which are larger than all the values in the lGRB models (see Table \ref{tab_summary}).
		The second difference is in the cocoon, which extends to smaller angles, with a coefficient $ f_c \approx 4 $ in the exponent (see Figure \ref{fig:sGRB_dists}d for the cocoon comparison between sGRB models). Both differences are expected since sGRB jets are more stable and share less of their energy with the cocoon and the jet-cocoon interface.
		
		Figure \ref{fig:sGRB_dists}e depicts the angular distribution of the energy-weighted proper-velocity. We find that the sGRB models follow Equation \ref{eq:utheta}, but having a more stable jet allows the terminal value of $ <u_{\infty,j}> $ at the core to be as high as $ u_{\infty,\mx} $. The lower mixing also results in a steeper power-law index $ p_u $. When considering $ \Eiso $ above a certain value $ u_i $ of proper-velocity, the behavior is similar to lGRBs, namely there is a cutoff at the angle that corresponds to $ <u_\infty> = u_i $ in the proper-velocity distribution.
		
		\subsubsection{Lower luminosity short GRBs}
		\label{sec:low_luminosity_sgrbs}
		
			The three sGRB models $ \SG_1, \SG_2 $ and $ \SG_3 $, which were motivated by the jet models that fit the afterglow of GW170817, include jets with isotropic equivalent luminosities $>3 \times 10^{51}~{\rm erg~s^{-1}}$, which are at the high end of the short GRB distribution \citep[see discussion at][and references therein]{Nakar2019}. The high isotopic equivalent luminosity possessed by these jets inhibits the growth of instabilities, thereby keeping the jets intact. Due to their narrow opening angle and relatively high luminosities these jets reach average head velocities of $ \sim 0.4c $ while propagating in the ejecta of $ \sim 0.05 \msun $. Subsequently the engine working time necessary for a successful breakout of these jets is a fraction of a second.

			The sGRB luminosity function indicates that while the above GW170817-like models maintain rather common observed sGRB energies, when considering the volumetric rate of sGRBs, bursts with $ \Liso \sim 10^{49}-10^{50}~{\rm erg~s^{-1}}$ are much more common than bursts $ \Liso \sim 10^{52}~{\rm erg~s^{-1}}$  \citep[e.g.][]{Guetta2005,Guetta2006,Nakar2006,Petrillo2013,Wanderman2015}. We therefore consider another model, $ \SG_4 $, with an identical setup to $ \SG_3 $, but with an isotropic equivalent luminosity of $ \Liso = 10^{49}~{\rm erg~s^{-1}} $.
			We find that model $ \SG_4 $ features an utterly different evolution than those of the other sGRB models. Its propagation is slower with an average head velocity of 0.22c before breakout, and its degree of mixing is rather high. Overall this model is similar to our lGRBs, owing to a similar ratio of jet to medium enthalpy density, $ \tilde{L} $.
			Figure \ref{fig:low_lumuinosity_sgrb} depicts the angular distribution of the isotropic equivalent energy of model $ \SG_4 $ when the jet head doubles the radius of the ejecta front. It is compared with that of model $ \A $ when the jet head reaches twice of the stellar radius. One can see that the distributions are very similar with high mixing that leads to a very shallow drop in the isotropic energy in the inner $ \sim 0.7\rad $. The two models differ in their cocoon energy due to the longer breakout time of the lGRB model which renders the cocoon more energetic.
		
			The breakout time (5.6s) of model $ \SG_4 $ implies that the engine of the jet has to work for at least $ \sim 4.5\s $ in order for the forward shock to exit the ejecta. This time is longer than the typical duration of sGRBs by about an order of magnitude.
			In general, the breakout time depends on the jet luminosity and ejecta mass in the Newtonian regime\footnote{Note however that model $ \SG_3 $ is approaching relativistic velocities before breakout.} \citep{Nakar2019}:
			\begin{equation}\label{eq:tb}
			t_b-t_d \propto \bigg(\frac{M_{ce}}{L_j}\bigg)^{1/3}~.
			\end{equation}
			We thus conclude that the typical duration of a sGRB is too short for jets with isotropic equivalent luminosities of $ \sim 10^{49}~{\rm erg~s^{-1}} $ to break out from a $ \sim 0.05
			\msun $ ejecta.
			A couple of possible intriguing solutions arise to that discrepancy:

			i) Total ejecta mass of $ 0.05 \msun $ is uncommon. Equation \ref{eq:tb} and the variance of $ L $ over orders of magnitude imply that the ejecta mass varies substantially as well.
			Previous numerical studies of NS mergers have shown that many configurations result in a substantially lower ejecta mass \citep[see discussion in][and references therein]{Nakar2019,Shibata2019}. Another possibility is that the ejecta is highly anisotropic, so that the effective mass that the the jet encounters is substantially lower. Either way, the lower ejecta mass in the direction of the jet reduces $ t_b $ to become comparable to those of models $ \SG_1, \SG_2 $ and $ \SG_3 $. In the less common cases where a lower luminosity jet propagates in a $ \sim 0.05 \msun $ ejecta, the jet is likely to be choked inside and do not produce a GRB.
			Interestingly, if the mass of the ejecta is indeed lower by an order of magnitude or more, it may imply that the r-process elements that are synthesized in the ejecta cannot account for the r-process abundance in the Universe \citep{Hotokezaka2018b}.

			Equivalently, it is also plausible that there is a correlation between the jet luminosity and the ejecta mass. This may happen, for example, if the relativistic and the sub-relativistic outflow components are both positively correlated with the mass of the accretion disc. Such a correlation dictates that $ t_b-t_d $ does not vary by much, with a typical value that is sufficient for most sGRB jets to break out.

			ii) The jet is at least weakly-magnetized. In the companion paper \citep{Gottlieb2020b} we show that weakly-magnetized jets are less prone to local hydrodynamic instabilities. Magnetized lGRB jets with a similar $ \tilde{L} $ to that in $ \SG_4 $, propagate $ \sim 3 $ times faster in the dense medium than their hydrodynamic counterparts. It then follows from Equation \ref{eq:tb} that the breakout of magnetized jets is equivalent to that of hydrodynamic jets which are 1.5 orders of magnitude more luminous.			Therefore, if the lower luminosity sGRB is weakly-magnetized, it is possible that it can break out in a rather typical sGRB duration even from a massive ejecta, as the one in GW170817.

		\begin{figure}
			\centering
			\includegraphics[scale=0.23]{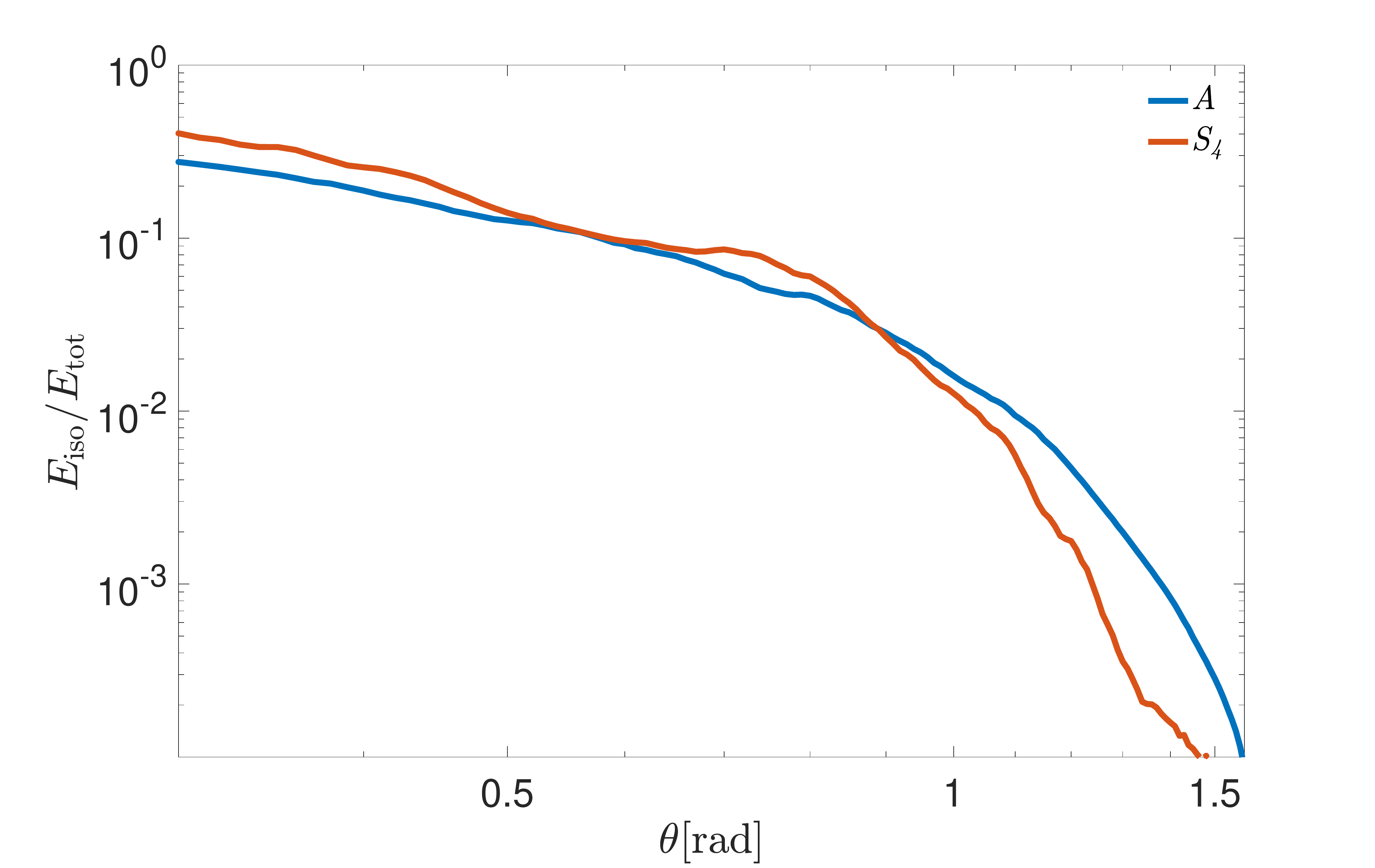}
			\caption[Low luminosity sGRB distribution]{
				The angular distribution of the isotropic equivalent energy of models $ \A $ (blue) and $ \SG_4 $ (red) when the jet head doubles the radius of the star and ejecta front, respectively. The distributions are normalized by the total energy in each system.
			}
			\label{fig:low_lumuinosity_sgrb}
		\end{figure}

		\subsection{A Comparison with Gaussian and power-law jet models}
		Post breakout GRB jets are often modeled as Gaussian structured jets with an energy distribution
		\begin{equation}
		\Eiso(\theta) = E_j{\rm exp}\bigg(-\frac{\theta}{2\theta_j}\bigg)^2~,
		\end{equation}
		where $ E_j $ and $ \theta_j $ are constants. This profile has been motivated by fits to 2D axisymmetric simulations (e.g. \citealt{Xie2018}). Here we examine whether such models can also be applied to jets that are found in 3D simulations. In Figure \ref{fig:gaussian} we show  Gaussian fits (dashed black lines) to the angular distributions of the isotropic equivalent energy of models $ \A $ and $ \SG_1 $, including the distribution of matter faster than certain values.
		Above we showed that the full energy distributions (blue and green) of all models can be approximated rather well by a flat core up to $\theta_j$ and a power-law of $\theta/\theta_j$ up to $ \theta_c $. As expected these distributions are incompatible with Gaussian fits.
		When considering the energy above a certain relativistic proper-velocity, the energy distribution does fall close to exponentially. Therefore it can be approximated by a Gaussian, although the fit is inadequate.
		When considering the angular distribution $u_\infty(\theta)$ (e.g., Figure \ref{fig:vtheta}), it is clear that a Gaussian is in tension with the power-law distributions that we found for both lGRBs and sGRBs.

		Recently, \citet{Lazzati2019} found that the angular energy and Lorentz factor profiles in axisymmetric 2D simulations can be fit by a double exponential function.
		As we show next in \S\ref{sec:2D}, 3D outflows are inherently different from 2D ones, and so are their angular structures. Therefore, as shown for the Gaussian fit, the double exponential function also falls too fast after the core to fit the JCI segment (which is absent in 2D, see \S\ref{sec:2d_dist}), and thus cannot describe 3D angular profiles.

		Another suggested model is a power-law distribution with a core. This model naturally reproduces the weighted average energy of the proper-velocity distribution with a power-law index $ p_u $ that mainly depends on the type of the system (lGRBs or sGRBs). In addition, a power-law model with a core also reproduces the isotropic equivalent energy distribution at $ \theta \lesssim \theta_c $ with a power-law index $ \delta $. Therefore we find that a power-law model is more consistent with our results than a Gaussian model. However, a single power model fits  only the jet and the JCI, but does not account for the cocoon distribution.
		
		\begin{figure}
			\centering
			\includegraphics[scale=0.23]{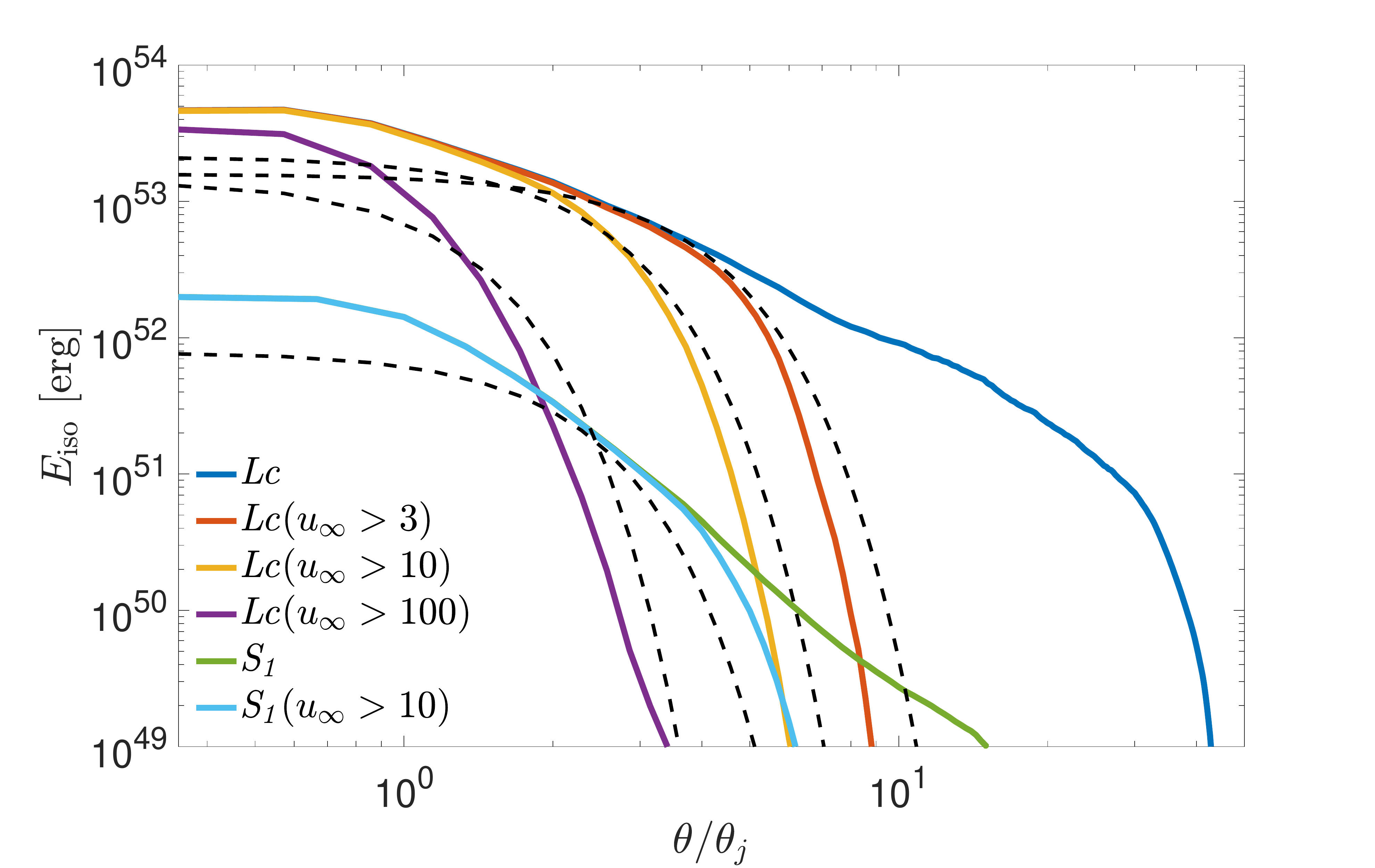}
			\caption[Gaussian fitting]{
				Gaussian fits (dashed black lines) to the angular distribution of the isotropic equivalent energy of models $ \A $ and $ \SG_1 $. Separated into the full distribution and the distributions above certain values of proper-velocities.
			}
			\label{fig:gaussian}
			
		\end{figure}
		
		\subsection{Comparison with 2D simulations}
		\label{sec:2D}
		
		Axisymmetric 2D jets feature an utterly different structure than 3D jets (see \citealt{Harrison2018} for a detailed comparison), which can be signified by two characteristics: (i) the structure of the jet head and (ii) the jet integrity.
		(i) While jets in 3D simulations are capable of wiggling around the accumulated stellar material on top of them, 2D jets are missing the extra dimension which allows such behavior, and thus they keep interacting with the heavy stellar material, denoted as ``the plug" \citep{Zhang2003}. In the aftermath of the jet evolution, the presence of the plug }{plays} a decisive role. The continuous interaction of the head with the plug deflects the relativistic jet material to large angles. As the jet is continuously losing material due to this interaction, it shrinks, and eventually dies off altogether (see \citealt{Gottlieb2018a}).
	(ii) We showed that instabilities develop in 3D models on the  $\hat{r}-\hat{\phi}$ plane.
	In axisymmetric simulations the $\hat{r}-\hat{\phi}$ plane does not exist and  instabilities cannot form on it. 
	The axisymmetric jets remain unperturbed on the $ \hat{r}-\hat{z} $ plane as well, and the jet-cocoon interface is absent, allowing the recollimation shocks to remain intact.
	The two differences can be seen in meridian slices of 2D (top) and 3D (bottom) post breakout jets, presented in Figure \ref{fig:2d3dmaps}. The 3D jet boundary is diffused into the cocoon and the JCI is clearly present between the white and black dashed lines, whereas the 2D jet remains unperturbed. In the 2D jet the plug is seen as the energetic component on top of the second recollimation shock in the top panel. It deflects jet material to large angles, leading to a formation of an energetic arc below the bow shock. The remaining parts of the system, i.e. the non-relativistic components are similar in 2D and 3D. We show that subsequently the 2D and 3D post-breakout distributions also exhibit substantial differences.
	
	\subsubsection{2D distributions}
	\label{sec:2d_dist}
	
	\begin{figure}
		\centering
		\includegraphics[scale=0.23]{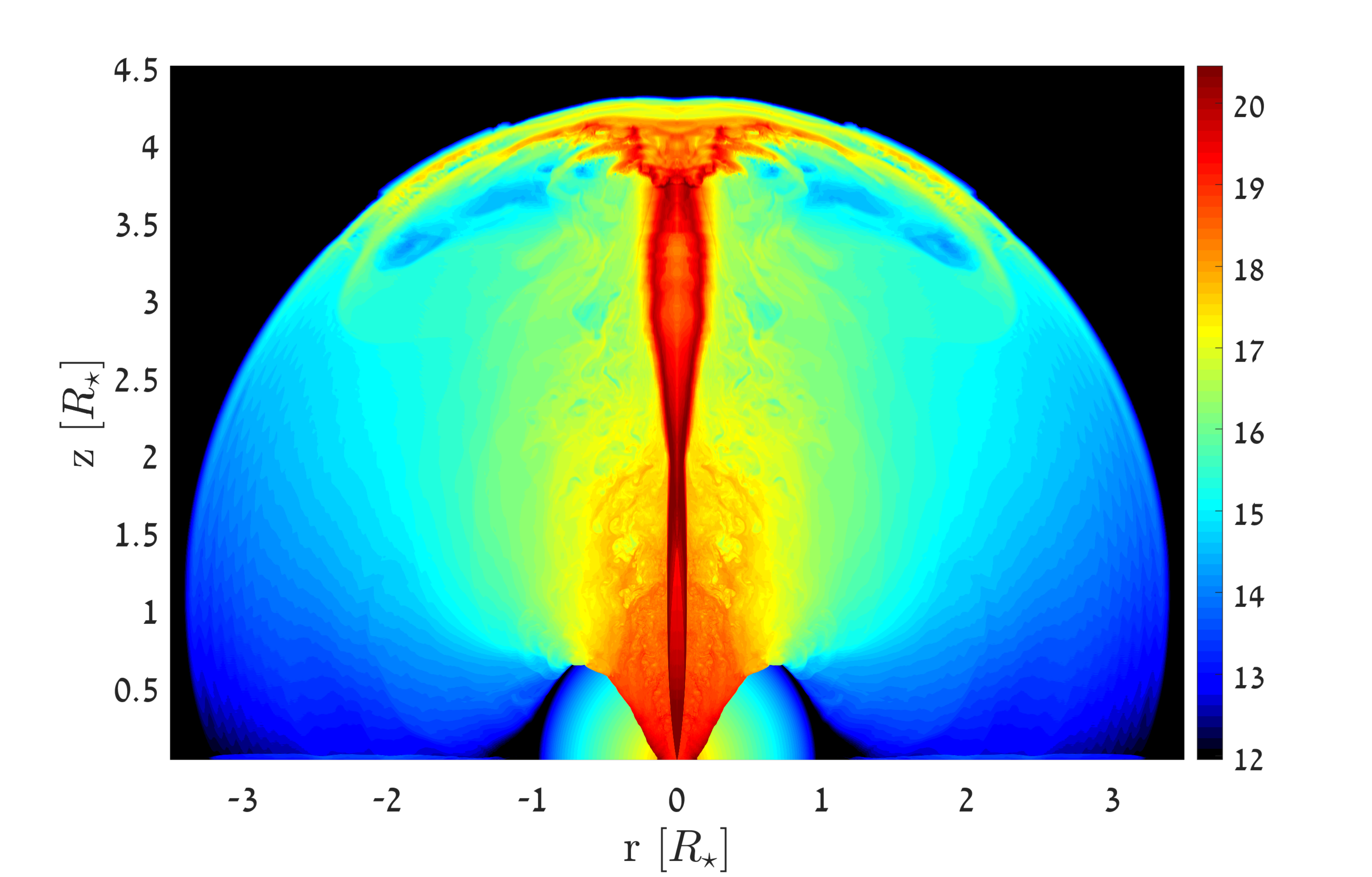}
		\includegraphics[scale=0.23]{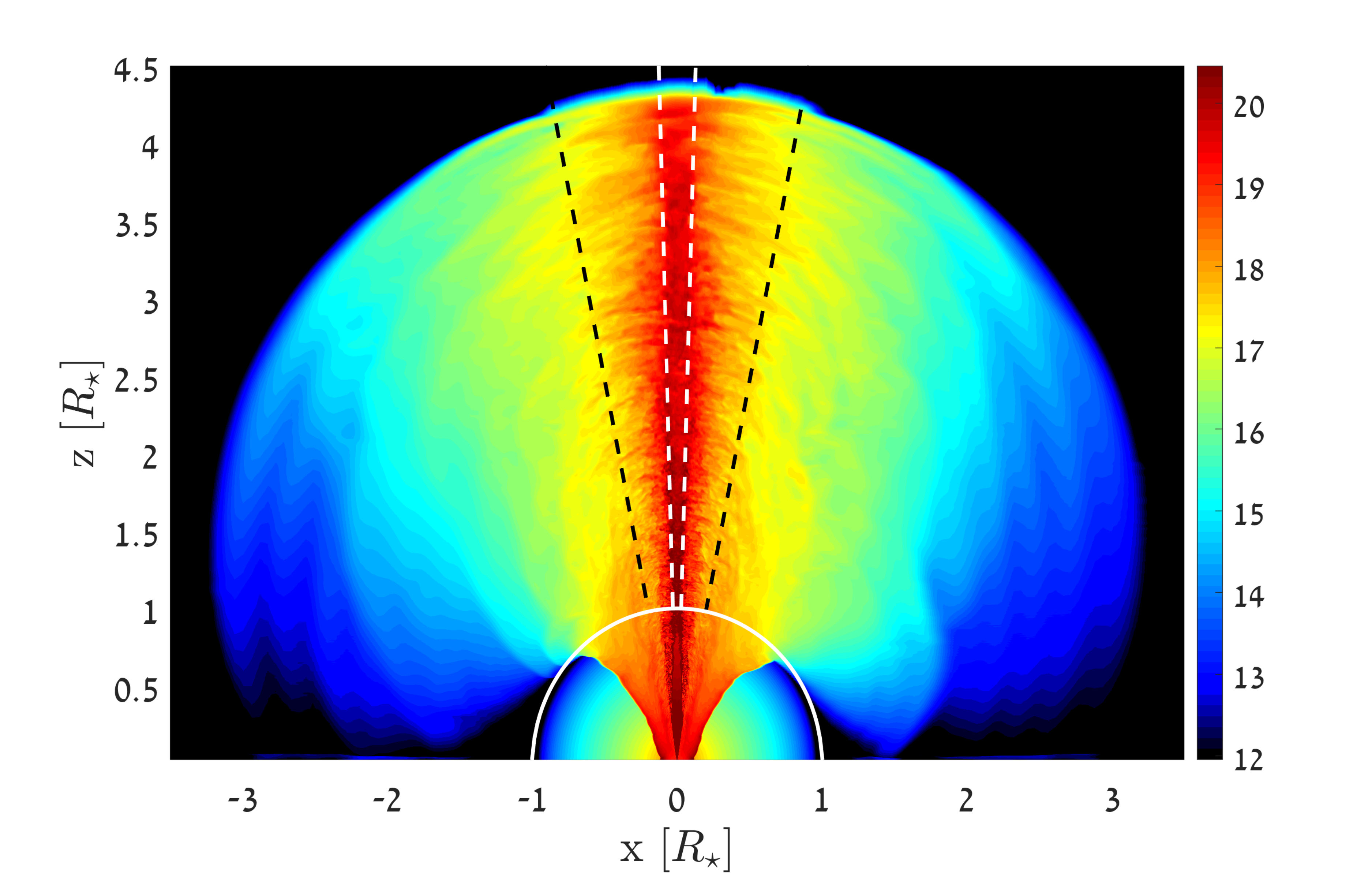}
		\caption[2D3D]{
			The logarithmic energy density $ [\erg~\cm^{-3}] $ units maps of model $ \G $ in 2D (top) and 3D (bottom). In the 3D model the continuous white line denotes the star, the dashed white lines delimit the jet core ($ \theta = \theta_j $), and the black lines delimit the jet-cocoon interface ($ \theta_c = 0.3\rad $). Beyond the dashed black lines begins the cocoon. 
		}
		\label{fig:2d3dmaps}
	\end{figure}
	
	\begin{figure}
		\centering
		\includegraphics[scale=0.23]{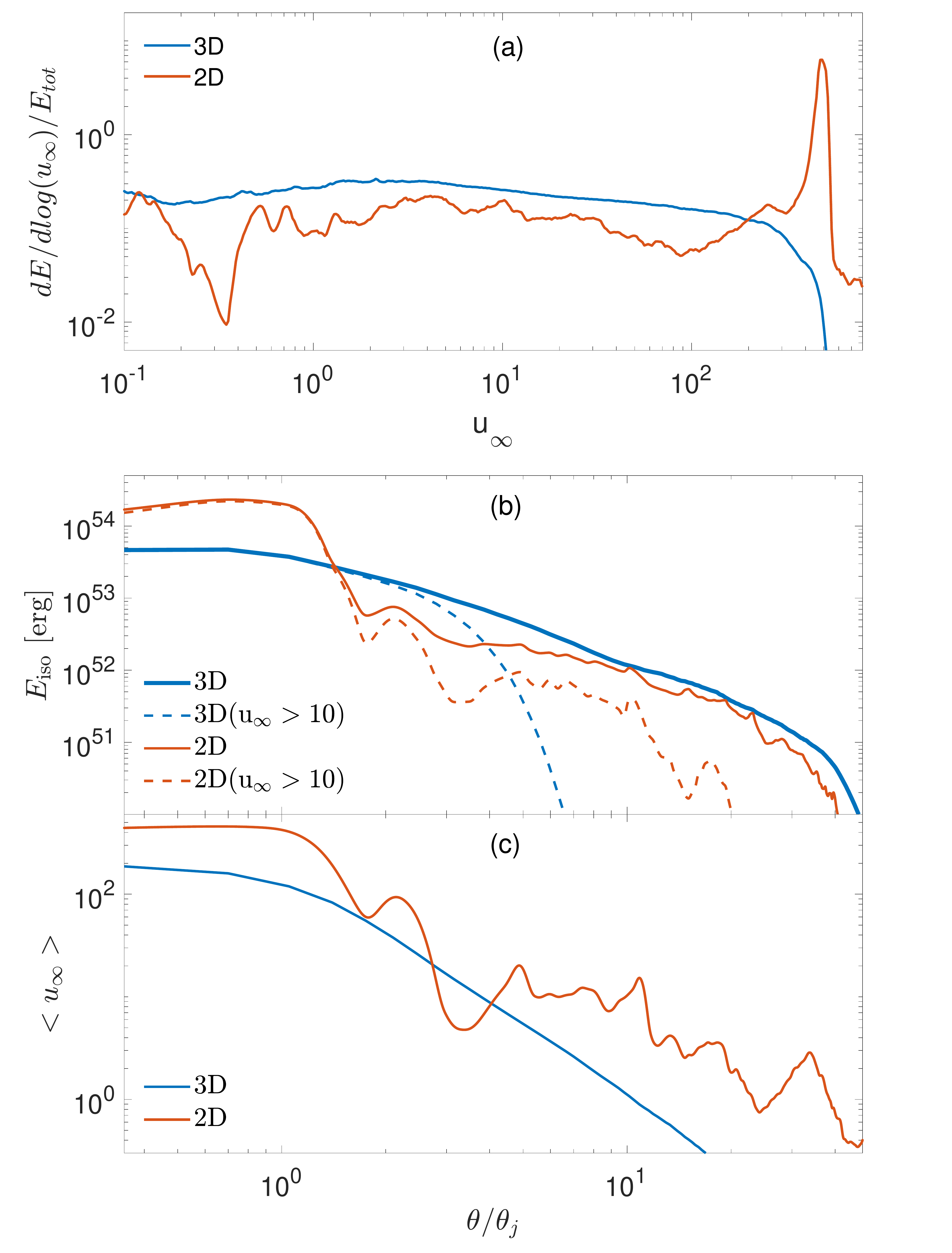}
		\caption[2D]{
			2D (red) vs. 3D (blue) distributions of matter that broke out from the star in the canonical model $ \A $ when the jet reaches $ 10R_\star $.
			Top: The energy distribution per a logarithmic scale of the terminal proper-velocity, normalized by the total energy of each curve, similar to Figure \ref{fig:Ehg}.
			Middle: The isotropic equivalent energy, similar to Figure \ref{fig:Etheta}.
			Bottom: The energy-weighted average of the proper-velocity, similar to Figure \ref{fig:vtheta}.
		}
		\label{fig:2D}
	\end{figure}
	
	Figure \ref{fig:2D} depicts a comparison of the distributions in 2D and 3D simulations of model $ \A $. Figure \ref{fig:2D}a demonstrates that the 2D jet retains its energy at $ u_\infty = u_{\infty,\mx} $ unlike the flat 3D distribution. In lower values of $ u_\infty $ the 2D mixing which originates only in the jet head (rather than both the head and the JCI in 3D) leads to a similar roughly flat distributions in 2D and 3D.
	
	The distinct energetic flat core in the 2D simulations can also be seen in the isotropic equivalent energy distribution (Figure \ref{fig:2D}b). This, of course, comes at the expanse of the JCI which is not really present in 2D, so there is an order of magnitude drop in the energy distribution outside the core.
	Another substantial difference between 2D and the 3D models lies in the energy distribution of matter that moves at velocities larger than given values. For example we show in dashed lines the energy at $ u_\infty > 10 $. The 3D model shows a sharp drop at $ \theta \approx 4\theta_j $, so that to a good approximation all the matter at $ \theta < 4\theta_j $ moves at $ u_\infty > 10 $, and all the matter at $ \theta > 4\theta_j $ moves at $ u_\infty < 10 $. In the 2D simulation on the other hand, even at $ \theta = 10\theta_j $ the matter with $ u_\infty > 10 $ has a comparable amount of energy to that with $ u_\infty < 10 $. This occurs due to the deflection of jet material by the plug. The 2D mildly-relativistic material in the cocoon ($ \theta \gtrsim \theta_c $) resembles that of the 3D.
	
	Figure \ref{fig:2D}c  depicts The energy-weighted average proper-velocity.
	The 2D distribution features $ u_{\infty,j} \approx u_{\infty,\mx} $, owing to to the absence of mixing at the jet core.
	At the edge of the 2D core there is a sharp drop followed by a bump at $\theta/\theta_j \approx 2$. The bump represents a large amount of energy in matter with low baryon contamination and is the signature of the jet material that was deflected sideways at the head by the plug. Similarly, one can see that at larger angles the 2-dimensional $ <u_\infty> $ is higher by almost an order of magnitude than its 3D counterpart, as a result of energetic arc that is formed by the plug.

	\section{Emission}
	\label{sec:emission}
	
	\subsection{Prompt}
	
	After its breakout, the jet accelerates and produces the prompt gamma-ray burst emission.
	The details of the origin of the prompt emission
	are still obscure and are not addressed here.
	Instead, we highlight some robust features that the mixing has on the emission, those are independent of the specifics of the emission process:
	In \S \ref{sec:lgrbs} we showed that due to the heavy baryon contamination in most hydrodynamic lGRB jets, the jet Lorentz factor upon breakout is relatively low, having $ u_\infty \ll u_{\infty,\mx} $\footnote{The low Lorentz factor in 3D hydro jets has already been seen in previous GRB jets' simulations, e.g. \citet{Zhang2003,Rossi2008,LopezCamara2013,LopezCamara2016,Harrison2018,Gottlieb2018a,Gottlieb2018b}.}. Over time the mixing decreases and $ u_\infty $ reaches typical values of GRBs. The temporal evolution of the Lorentz factor implies the following:
	(i) The observed light curve and spectrum should show some kind of evolution, particularly in lGRBs.
	(ii) Some jets that break out may not live long enough to reach Lorentz factors that are necessary to produce a GRB (see e.g. the early time distribution in Figure \ref{fig:Ehg}b).
	It is possible that the emerging outflow of such jets resemble the one obtained for intermittently launched hydrodynamic jets.
	\citet{Gottlieb2020a} recently showed that while engines of GRBs are considered to be intermittent, 3D simulations of variable hydrodynamic jets suggest that such jets are subject to heavy baryon entrainment. Therefore their terminal Lorentz factor is too low to produce a $ \gamma $-ray emission.
	The lack of evidence for a temporal evolution in the prompt light curve, as well as the difficulties in producing $ \gamma $-rays due to the mixing in hydrodynamic jets\footnote{The mixing in modulated jets is likely to be of a different origin.} may challenge the hydrodynamic jet model for lGRBs. Some level of magnetization in the jet can inhibits the mixing and may solve these issues \citep{Gottlieb2020b}. 
	
	The effect of the mixing on a particular type of emission mechanism was previously considered in \citet{Gottlieb2019b}, who presented analytic model and detailed numerical calculations of the photospheric emission using numerical simulations.
	They showed that the radiative efficiency at the photosphere is inevitably high as long as $ u_{\infty} \gtrsim 100 $, and that the mixing plays an important role in shaping the photospheric emission. The effects of the mixing can be summarized as follow:
	(i) It reduces the radiative efficiency by lowering the average terminal proper-velocity in the jet, ($u_\infty$).
	(ii) It results in high variability in the efficiency, similar to the observed variability timescales in the prompt signal.
	(iii) Different elements break out from the star with different $ u_\infty $, naturally producing internal shocks below the photosphere.
	(iv) The mixing alters the off-axis emission.
	(v) The change in the degree of mixing with time leads a temporal evolution of the efficiency.
	The analysis of \citet{Gottlieb2019b} considered models $ \A, \D, \h, \I $ and $ \J $ in which the mixing diminishes with time, and the efficiency increases. In very powerful and narrow jets the evolution of the mixing is reversed, as fresh material exits the star increasingly mixed. This would generate a photospheric emission with a reversed temporal evolution of the efficiency.
	
	\subsection{Afterglow}
	
	At long times the interaction of the jet-cocoon outflow with the interstellar medium (ISM) generates the afterglow emission via synchrotron radiation. The variance in the angular distribution of the outflow, most notably between lGRBs and sGRBs, entails different afterglow signatures.
	The afterglow light curve is shaped by the viewing angle $ \theta_{\rm obs} $ and by the mixing, which determines the power-law index $ \delta $ of the isotropic equivalent energy in the JCI.
	In outflows with weak mixing where $ \delta > 2 $, as we find in the sGRB setups $ \SG_1, \SG_2, \SG_3 $ and in a few of our lGRB setups, most of the outflow energy resides in the jet core. The afterglow from such jets is similar to that of a top-hat jet for any observer who is within the opening angle of the jet-core, namely at $\theta_{\rm obs}<\theta_j$. The afterglow seen by an observer at $\theta_{\rm obs}>\theta_j$ peaks at late time and the light curve during the rise depends mostly on whether the line of sight intersects with the JCI (i.e., $\theta_j<\theta_{\rm obs}<\theta_c$) or not (i.e., $\theta_c<\theta_{\rm obs}$). In the former case the rising phase is more complex, with several different possibilities: a hump before the main peak,  an early peak followed by a shallow decay, or a double peak \citep{Beniamini2020}. The early rise to a hump/peak is generated by the JCI material that moves towards the observer while the second/main peak is generated by the jet core. In a case where $\theta_c<\theta_{\rm obs}$, the light curve starts with a sharp rise and then followed by a shallower rise to the peak (presumably similar in shape to the afterglow of GW170817, where only the shallow rise phase was detected). In both cases the peak is often seen when the jet core decelerates enough so its beamed emission includes the observer. The light curve at later times, after the peak is similar to the one seen from a top-hat jet. As was demonstrated by \citet{Gottlieb2019a}, the time and the flux of the peak in these cases are described by the simple analytic formula derived for a top-hat jet seen off-axis \citep[e.g.][]{Nakar2002,Granot2002}.
	
	When the mixing is strong such that $ \delta < 2 $, as we find in most of our lGRB models and in $ \SG_4 $, the JCI possesses most of the outflow energy. There are two main possibilities for the afterglow shape that depend on the location of the observer.
	(i) $ \theta_{\rm obs} < \theta_j $: The observer sees at first a decay that is similar to that of an on-axis top-hat jet with a break at the point where the decelerating jet-core reaches a Lorentz factor $ \Gamma \approx \theta_j^{-1} $. However, since $ \delta < 2 $ the total energy that is within the observer view increases with time also when $ \Gamma < \theta_j^{-1} $,  therefore the light curve profile after the break is shallower than in the case of a top-hat jet. Once the blast wave decelerates enough so most of the energy of the JCI is observed, the power-law of the break gradually converges to the  post jet-break light curve of a top-hat jet \citep{Sari1999c}. Therefore, for observers who face the jet we expect another light curve segment which exhibits a gradual transition from the on-axis emission to the steep post jet break power-law.
	(ii) $ \theta_{\rm obs}>\theta_j $: The observer sees a rise, which originates in the JCI to a peak, which is followed by a shallow decline up to the time where the jet core becomes visible. At this point the decline steepens and the light curve coincides with the one seen by an observer that is along the jet axis. The jet core in this case plays only a secondary role and therefore, the analytic formula for the time and flux of the peak that was derived for an off-axis top-hat jet are not applicable.
	
	The aforementioned results are applicable to most of our lGRB models, and thus imply that sGRB and lGRB afterglows may be inherently different. However, there is no clear observational evidence that this is indeed the case \citep{Nakar2007,Berger2014}. This may have several different reasons. One is that the difference between the light curves predicted for different values of $\delta$ is more obvious for observers at $\theta>\theta_j$. However,
	all GRBs, long and short, were presumably observed at $ \theta_{\rm obs} < \theta_j$, with the exception of GW170817. Thus, it is possible that the signature of events with $\delta<2$ was missed, especially since no one looked for this signature and it may be hard to detect if $\delta$ is not much smaller than 2, as we find in quite a few of our lGRB models. Another possible reason for why the signature of $\delta<2$ was not observed, may be simply that lGRB jets are not hydrodynamic and therefore there is much less mixing and $\delta>2$ in lGRBs as well \citep{Gottlieb2020b}.
	Finally, the reason for the absence of evidence of differences between lGRBs and sGRBs may also lie in the fact that the jets possess a stabilizing magnetic component. In a companion paper \citep{Gottlieb2020b} we show that magnetized jets feature a similar post-breakout structure to hydrodynamic sGRBs. Thus, if GRB jets are magnetized inside the dense medium, the afterglows of sGRBs and lGRBs will not be so different after all.
	
	\section{Summary}
	\label{sec:discussion}
	
	We present a study of the structure of GRB outflows based on a set of 3D simulations of relativistic hydrodynamic GRB jets that interact with the dense media that surround the launching sites. We find that all simulations exhibit Rayleigh-Taylor fingers that emerge from the cocoon into the jet at an early stage. The fingers grow with time and lead to an intense mixing between the jet and the cocoon. The mixing disrupts the coherent structure of the jet and gives rise to a transition layer denoted as the jet-cocoon interface (JCI).
	The mixing differs between lGRBs and sGRBs. 
	In lGRBs after the breakout the mixing in the star evolves over a timescale of $ T \approx t_b $ to a uniform distribution of energy in the proper-velocity logarithmic space. In sGRB models $ \SG_1, \SG_2 $ and $ \SG_3 $ (but not in $ \SG_4 $) the mixing is less important and the jet maintains about half of the injected energy in the core.
	This phenomenon has profound implications on the evolution of the jet structure after breakout as well as on the prompt and afterglow emission signatures.
	
	We find that almost all the energy that is launched during the jet propagation through the medium (i.e. prior to the breakout) is deposited in the cocoon. After the breakout the cocoon's energy remains constant and freshly launched energy is deposited in the jet core and in the JCI.
	The JCI, which bridges  the jet-core and the cocoon, allows a smooth transition between the two components. The energy is divided between the jet and the JCI,  where the energy fraction in the jet core, $ \lambda $, reflects the stability of the jet and dictates the shape of the JCI. Typically lGRB jets are less stable and maintain smaller values than sGRBs of $ \lambda \lesssim 0.2 $. Subsequently, their JCIs contain more energy distributed over larger angles and dropping slower with $\theta$. Short GRB jets, which propagate through much thinner media and are typically more stable, maintain larger values of $ \lambda \approx 0.4 $ and feature steeper JCI energy distributions.
	
	However, sGRBs with luminosities $ \Liso \lesssim 10^{50} {\rm erg~s^{-1}} $ are considered to be more common across the universe. If they are accompanied by a massive ejecta, their evolution is similar to that of lGRBs, owing to their similar jet to ejecta enthalpy ratio. Such jets cannot break out from a massive ejecta during the typical engine times of sGRBs. This result may rise several interesting possibilities regarding the nature of the jet, ejecta and their interrelation.
	
	We provide a full analysis of the structure of a variety of lGRBs and sGRBs. We find that in both types of GRBs the distribution of isotropic equivalent energy can be approximated by an angular structure with a flat core of the jet, followed by a power-law distribution of the JCI  with a power-law index $\delta $, that is set by the value of $ \lambda $. The cocoon structure is more complex, including both a radial and an angular structure. The angular distribution of its isotopic energy energy can be approximated by an exponent. In addition to the energy distribution, we find that the distribution of the outflow proper-velocity can also be approximated by a flat core at the jet and a power-law at larger angles. These profiles are similar to the popular power-law jet models, with the exception that a power-law distribution does not account for the cocoon and is therefore valid only at small angles (in our models $\theta \lesssim 0.3$ rad). The structures that we find cannot be approximated well by the frequently used Gaussian jet model.
	
	We compare the results of 3D and 2D simulations showing that 2D simulations both lack the feature of mixing, and are subject to numerical artifacts. As a result, 2D models yield considerably different structures than 3D ones. We therefore conclude that 2D simulations of the interaction between GRB jets and the surrounding star or merger ejecta are of limited accuracy and in particular cannot be used for studying the structure of the jets before or after they break out.

	The mixing plays a crucial role in shaping the prompt emission.
	\citet{Gottlieb2019b} demonstrated the strong effect that mixing has on the photospheric prompt emission. However,
	some mixing effects are relevant for any emission mechanism. For example, the evolution in the mixing in lGRBs should lead to a temporal evolution in the prompt light curve. The fact that such evolution is not seen is in tension with  hydrodynamic jets being the source of GRBs.
	The afterglow is also affected by the structure of the outflow.  When $ \delta < 2 $, as we find in most of our lGRB models, most of the jet energy resides in the JCI. The result is that the jet break seen by an observer within the jet opening angle takes a slightly different form than the one obtained for the typical top-hat jet. So far such signature was not identified, although it may be hard to detect. If the observer line-of-sight is outside of the jet core, the peak of the afterglow light curve is dominated by the JCI rather than by the jet core.
	This characterization may hold promise in inferring some of the fundamental jet-medium characteristics from afterglow observations.

	Finally, it is noteworthy to mention that GRB jets can also take other forms.
	In a companion paper \citep{Gottlieb2020b} we show that magnetic fields, even if they are subdominant, can stabilize the jet boundary and reduce the mixing considerably. This allows the jet to maintain a larger fraction of its original energy, similar to the hydrodynamic sGRB jets in models $ \SG_1, \SG_2 $ and $ \SG_3 $. Consequently, the structure of magnetic jets after breakout differs substantially from that of hydrodynamic jets. As a result, their emission imprint is also expected to show other characteristics, and in principle may allow observations to infer the nature of GRB jets and determine whether they are magnetic or hydrodynamic at their base.
	A different type of jets is choked jets. Previous works have suggested that the majority of jets of collapsing stars never break out as their engine activity time is not sufficiently long to push them out of the star \citep{Mazzali2008,Bromberg2011,Sobacchi2017}.
	In such cases the cocoon is the only component that breaks out of the star, spreading quasi-spherically and forms an entirely different structure from a relativistic jet that breaks out. It is hence also interesting to study in detail the emission properties from such systems and how they are modified by the mixing.
	
	\section*{Acknowledgements}
	This research is partially supported by an ERC grant (JetNS) and an ISF grant (OG and EN).
	
	\bibliographystyle{mnras}
	\bibliography{Hydro_structure}
	
	\appendix
	\section{Convergence Tests}
	\label{sec:convergence}
	
	We verify that our results are independent of the numerical setup. We showed that the mixing is determined inside the star and does not change after breakout. Therefore, we test convergence for the mixing upon breakout, by carrying out two comparisons. The first is of the original resolution with a higher one. The higher resolution grid includes three patches on the $ \hat{x}-\hat{y} $ plane and one patch on the $ \hat{z} $-axis. On $ \hat{x}-\hat{y} $ the inner patch now includes 400 uniform cells in the inner $ r = 2.5\times 10^9\cm $. This increases the resolution inside the jet by approximately a factor of 2 and much better resolves the jet-cocoon interface. The  outer patches have 160 logarithmic cells until $ R_\star $, and therefore also improves the resolution in the cocoon. On the $ \hat{z} $-axis we employ 1000 uniform cells from $ z_{beg} $ to $ R_\star $.
	
	The second comparison is of the size of the nozzle.
	The mixing originates in the RT fingers that penetrate through the jet-cocoon interface before reaching the jet. In our simulation we find that the nature of the mixing is highly sensitive to the width of the jet upon injection. Jets that are injected with a wide nozzle are less vulnerable to mixing and are more stable compared to jets with smaller nozzle. Since the jets are generated in the vicinity of a compact object, their initial typical size is expected to be small. Therefore, one must verify that their nozzle is small enough to simulate their evolution properly.
	We carry out a simulation in which we inject a jet with half the size of the nozzle in our original simulation, and set $ z_{\rm{beg}} = r_{\rm{noz}}/\theta_{j,0} $ accordingly. We use 480 uniform cells in the inner $ 1.5\times 10^{9}\cm $, and 200 logarithmic cells that stretch outside of it until $ R_\star $. On the $ \hat{z} $-axis we use 1000 uniform cells from $ z_{beg} $ to $ R_\star $.
	
\begin{figure}
		\centering
		\includegraphics[scale=0.23]{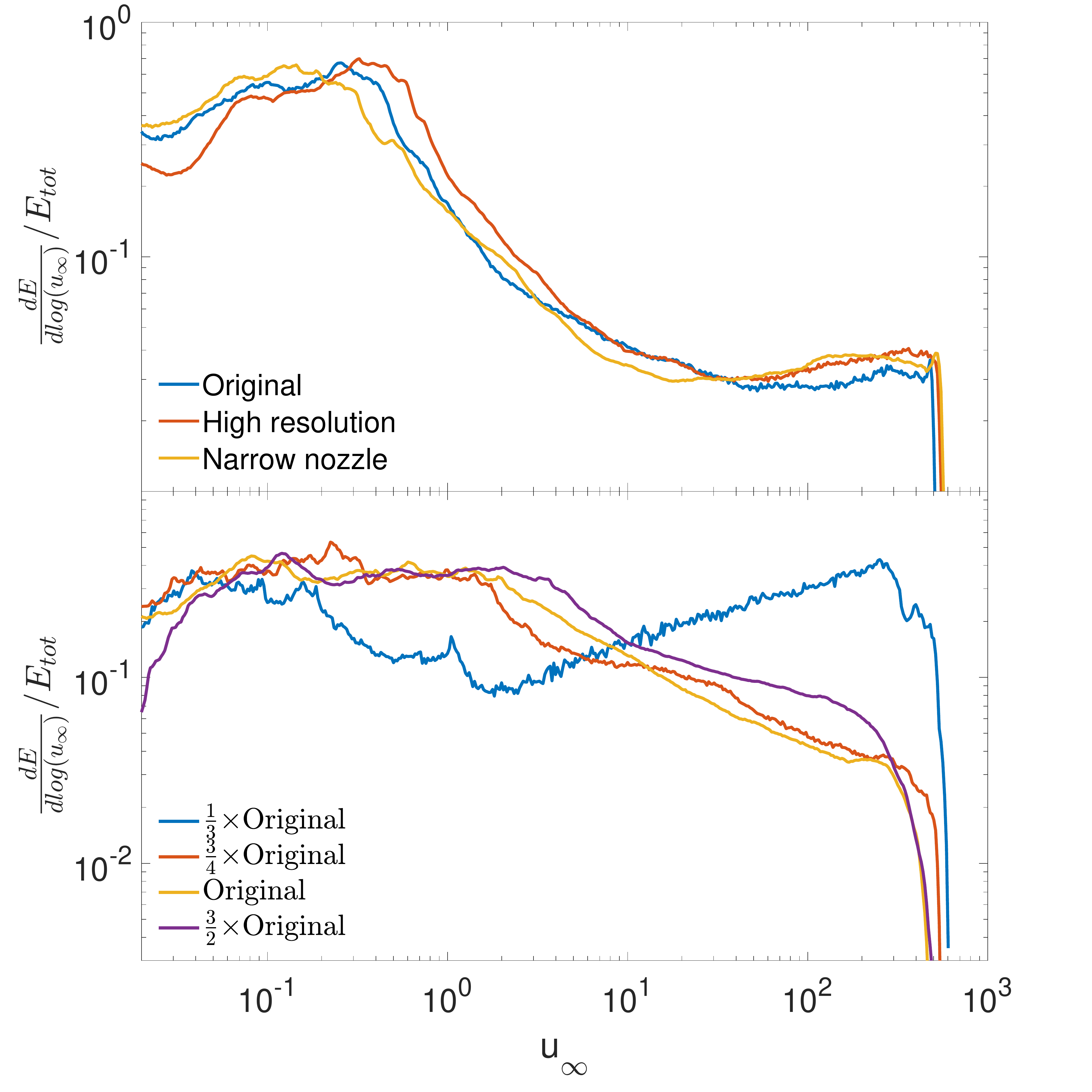}
		\caption[Afterglow]{
			Convergence tests in the logarithmic energy distribution as a function of $ u_\infty $ for the canonical model $ \A $. Top: Tests for resolution and nozzle size inside the star, above the collimation shock. Bottom: Tests for the post-breakout structure outside the star with four different resolutions compared to the original simulation. The distribution are taken when the jet head reaches $ 4R_\star $.
		}
		\label{fig:convergence}
	\end{figure}

	In the upper panel of Figure \ref{fig:convergence} we show that neither increasing the resolution nor reducing the size of the nozzle affects the mixing inside the star with all three curves are compatible with each other to a high degree. We stress that increasing the size of the nozzle above our original value leads to less mixing.
	One difference between the models is the breakout time. In the original resolution the breakout time is 20s, regardless of the nozzle size. Increasing the resolution shortens the breakout time, in our high resolution simulation test the breakout time is reduced to 15s. However, in this work we are interested in the distribution and the structure of the jet which are consistent with each other in all tests.
	
	We also verify that the post-breakout structure of the outflow remains similar at different resolutions. For this purpose we perform additional three simulations with the same physical setup of simulation $ \A $, but differ in their resolution. Specifically, $ \frac{1}{3} $, $ \frac{3}{4} $ and $ \frac{3}{2} $ the resolution of our original simulation, on both axes.
	In the lower panel of Figure \ref{fig:convergence} we show the resulting distributions when the jet reaches $ 4R_\star $. One can see that the behavior of the $ \frac{3}{4} $, $ \frac{3}{2} $ and original resolution are similar to each other with no particular trend between the three. The main difference between the simulations is a factor of two along the jet-cocoon interface.
	In the lowest resolution simulation mixing does not form, leading to a stable jet.
	
	\label{lastpage}
\end{document}